\documentclass[a4paper,12pt,english]{article}

\usepackage{amsfonts,bm,amssymb,euscript,array,babel,cite}
\usepackage{mathtools}
\usepackage{tikz-cd}
\usepackage{amsmath}
\usepackage{hhline}
\usepackage[amsmath]{empheq}
\usepackage{hyperref}
\usepackage{cancel}
\usepackage{mathrsfs}
\usepackage{pdfsync}
\usepackage{tikz}

\newcommand\ringring[1]{%
	{
		\mathop{\kern0pt #1}\limits^{
			\vbox to-1.85ex{
				\kern-2ex 
				\hbox to -1pt{\hss\normalfont\kern.1em \r{}\kern-.45em \r{}\hss}%
				\vss 
			}
		}
	}
}

\newcommand\ringringring[1]{%
	{
		\mathop{\kern0pt #1}\limits^{
			\vbox to-1.85ex{
				\kern-2ex 
				\hbox to -1pt{\hss\normalfont\kern.1em \r{}\kern-.45em\r{}\kern-.45em \r{}\hss}%
				\vss 
			}
		}
	}
}

\makeatletter
\newcommand{\doublewidetilde}[1]{{%
  \mathpalette\double@widetilde{#1}%
}}
\newcommand{\double@widetilde}[2]{%
  \sbox\z@{$\m@th#1\widetilde{#2}$}%
  \ht\z@=.9\ht\z@
  \widetilde{\box\z@}%
}
\makeatother

\newsavebox{\mysaveboxM}
\newsavebox{\mysaveboxT}

\makeatletter

\newcommand{\dd}{\mathrm{d}}

\newcommand{\w}{\wedge}

\newcommand{\be}{\begin{equation}}
\newcommand{\ee}{\end{equation}}
\newcommand{\sfrac}[2]{{\textstyle\frac{#1}{#2}}}
\def\nn{\nonumber}
\def \bea{\begin{eqnarray}} 
\def\eea{\end{eqnarray}}
\def\bse{\begin{subequations}}	
\def\ese{\end{subequations}}
\def\bal{\begin{align}} 
\def\eal{\end{align}}
\newcommand{\mf}{\mathfrak}
\def\mc{\mathcal}

\def\bi{\begin{itemize}} 
\def\ei{\end{itemize}}

\makeatother

\newtheorem{theorem}[equation]{Theorem}

\newtheorem{lemma}[equation]{Lemma}
\newtheorem{prop}[equation]{Proposition}

\def\a{\alpha} \def\b{\beta}   \def\d{\delta} 
\def\e{\epsilon} 
  \def\h{\eta} \def\k{\kappa}
\def\l{\lambda}  
 \def\o{\omega}   
\def\s{\sigma}   \def\th{\theta}
  \def\z{\zeta}
\def\O{\Omega}

\def\R{{\mathbb R}}  \def\N{{\mathbb N}}  
 
 \def\Z{{\mathbb Z}} 

\def\one{\mbox{1 \kern-.59em {\rm l}}}

\numberwithin{equation}{section}

\begin{document}

\makeatother
\parindent=0cm
\renewcommand{\title}[1]{\vspace{10mm}\noindent{\Large{\bf #1}}\vspace{8mm}} \newcommand{\authors}[1]{\noindent{\large #1}\vspace{5mm}} \newcommand{\address}[1]{{\itshape #1\vspace{2mm}}}

\begin{titlepage}

\begin{flushright}
 \today
\end{flushright}

\begin{center}

\title{ {\Large {A unified approach to standard and exotic \\[4pt]dualizations through graded geometry}}}

  \authors{\large Athanasios {Chatzistavrakidis}{$^{\a,}$}{\footnote{Athanasios.Chatzistavrakidis@irb.hr}}, Georgios Karagiannis{$^{\a,}$}{\footnote{Georgios.Karagiannis@irb.hr}}, Peter Schupp{$^{\b,}$}{\footnote{p.schupp@jacobs-university.de}}
    }
 
 \vskip 3mm
 
  \address{ $^{\a}$ Division of Theoretical Physics, Rudjer Bo\v skovi\'c Institute \\ Bijeni\v cka 54, 10000 Zagreb, Croatia \\ \vskip 3mm $^{\b}$ Department of Physics and Earth Sciences, Jacobs University,\\  28759 Bremen, Germany 
 
 }

\vskip 2cm

\begin{abstract}

Gauge theories can often be formulated in different but physically equivalent ways, a concept referred to as duality. Using a formalism based on graded geometry, we provide a unified treatment of all parent theories for different types of standard and exotic dualizations. Our approach is based on treating tensor fields as functions of a certain degree on graded supermanifolds equipped with a suitable number of odd coordinates.
We present a universal two-parameter first order action for standard and exotic electric/magnetic dualizations and prove in full generality that it yields two dual second order theories with the desired field content and dynamics. Upon choice of parameters, the parent theory reproduces (i) the standard and exotic duals for $p$-forms and (ii) the standard and double duals for $(p,1)$ bipartite tensor fields, such as the linearized graviton and the Curtright field. Moreover, we discuss how deformations related to codimension-1 branes are included in the parent theory. 

\end{abstract}

\end{center}

\vskip 2cm

\end{titlepage}

\setcounter{footnote}{0}
\tableofcontents


\section{Introduction and main results}
\label{sec1}

\subsubsection*{Motivation}

Electric/magnetic duality has its origins in Maxwell theory, where under a suitable exchange of the electric and magnetic fields the field equations remain the same in vacuum or when both electric and magnetic sources are included. In terms of the Faraday tensor, electromagnetic duality corresponds to the exchange of the tensor itself with its Hodge dual, which translates at the level of gauge potentials to the exchange of the Maxwell 1-form potential with a dual 1-form potential. Thus one obtains covariant formulations in terms of different gauge fields, which however lead to the same physical theory. 

This concept of duality is readily generalized for theories containing differential forms of higher degree in arbitrary dimensions. If an Abelian gauge theory for a $p$-form potential in $D$ dimensions is considered, then it is not difficult to see that it is dual to a theory formulated in terms of a $(D-p-2)$-form potential. Although the two fields do not have the same number of components, they describe the same  physical degrees of freedom under the little group. As in Maxwell theory, the corresponding field strengths are Hodge dual to each other. 

The above simple logic based on Hodge duality seems to indicate that there are two equivalent covariant descriptions of the same physical field. However, this is not precisely true, in the sense that there may exist more than two such descriptions. 
Even in the simple case of $p$-forms, it has been shown that exotic dualities exist too \cite{Hull,Bekaert:2002dt,deMedeirosHull1,Boulanger1,Boulanger2}, where a $p$-form has a dual field being a mixed symmetry tensor field of type $(D-2,p)$. Mixed symmetry tensor fields are generalizations of differential forms based on higher complexes and their generalized cohomology \cite{DuboisViolette1,DuboisViolette2}.  Their components in local coordinates contain more than one sets of antisymmetrized indices. We call them $N$-partite or multipartite tensors then. Simple examples of bipartite tensor fields are the metric tensor and the Curtright field \cite{Curtright:1980yk}, being $(1,1)$ and $(2,1)$ mixed symmetry fields respectively, while multipartite ones are  motivated from the study of higher spin theories. 

Mixed symmetry tensor fields are interesting also from the point of view of string and M-theory. They are present in abundance in the $E_{11}$ approach \cite{West:2001as} as brane charges \cite{Cook:2004er,West:2004kb}. This also indicates that, string-theoretically, low codimension branes (such as exotic, solitonic branes, domain walls and spacetime filling ones) couple electrically to mixed symmetry tensor fields \cite{Bergshoeff:2011zk,Bergshoeff:2010xc,deBoer:2012ma,Bergshoeff:2015cba,Chatzistavrakidis:2013jqa,Chatzistavrakidis:2014sua,Bakhmatov:2017les,Otsuki:2019owg,Fernandez-Melgarejo:2018yxq}, which can be exotic electromagnetic duals of the graviton or the Kalb-Ramond field. As such they are generalizations of magnetic monopoles both in higher dimensions and for (linearized) gravity.

Then it is natural to ask whether dualities for such tensor fields exist too. The answer is affirmative, at least at the linearized level, and indeed the graviton can be dualized to another field of type $(D-3,1)$ \cite{Hull,West:2001as}. Furthermore, tensors such as the Curtright field may have more than one distinct dual fields of the same type. This logic extends even further, in that mixed symmetry tensors also have double, or multiple for that matter, dual fields; an example would be a double dual graviton of type $(D-3,D-3)$ \cite{Hull,Boulanger2}. Finally, the extreme extension would be to find an infinite number of ways to represent a given degree of freedom in terms of dual covariant gauge theories, which was proven true in Ref. \cite{Boulanger:2015mka} in terms of higher and higher exotic dualizations. 

A natural way to relate dual theories is in terms of a parent or master action functional. This is typically a first order in derivatives action containing two independent $GL(D)$-reducible fields such that integrating out each of them leads to the two dual second order theories. Parent actions are simple to construct and analyze for differential forms and their standard duals. However, for exotic duality and for mixed symmetry tensor fields a number of non-trivial tricks are required \cite{Bergshoeff,Boulanger2003,Bergshoeff:2016gub}. 
Constructing the parent action in components may involve partial integrations in advance, while the analysis leads to non-trivial algebraic cancellations of unwanted terms, often after cumbersome calculations. Moreover, the starting point appears different for each type of field. One of the goals of the present paper is to unify these different parent actions in a single geometric one and prove in general that it leads to the desired dual descriptions for all types of dualizations.

To achieve the above goal, we utilize the framework of graded geometry (see e.g. \cite{Severa2001,dee1}, and \cite{Qiu:2011qr} for a review.) There are several reasons why graded geometry appears useful in this context. The most direct one is that mixed symmetry tensor fields are simply functions of a certain degree on a suitable graded supermanifold with a number of odd coordinates associated to it \cite{Chatzistavrakidis1} (see also \cite{Bruce}). One may then already ask a more primitive question than how to construct parent actions. How can one write kinetic, mass and higher derivative interaction terms for mixed symmetry tensor fields in a purely geometric and coordinate free way? 
Recall that this is certainly well known for differential forms, using the Hodge star operator and the wedge product. However, the situation is not as simple for more general cases, even for linearized general relativity let alone Lovelock gravity with higher derivative terms. 

Part of this question was addressed in Refs. \cite{deMedeirosHull1,Bekaert:2003az,deMedeirosHull2,deMedeiros:2003qel}, based on a formalism directly generalizing differential forms. On the other hand, higher derivative interaction terms were set in a graded geometric language in \cite{Chatzistavrakidis1}, where also all known kinetic Lagrangians for scalar fields, differential forms and bipartite mixed symmetry tensors were written in this formalism. This resulted in a general description of all Galileon-type higher derivative interactions for any collection of multiple species of such fields, which turned out to be very simple in geometric terms. 

\subsubsection*{Summary of results and outline} 

According to the above discussion and goals, 
we begin in Section \ref{sec2} by setting the graded geometric stage for the study of bipartite tensors. We discuss how they can be associated to functions of a certain degree on a suitable graded supermanifold equipped with sets of degree~1, odd coordinates. We highlight the definition and properties of a Hodge star operator~$\star$, which maps $(p,q)$ bipartite tensors to $(D-p,D-q)$ ones. This operator is instrumental in defining an inner product that allows us to construct geometric action functionals for such fields using Berezin integration in all sets of odd coordinates. In addition, we discuss the intrinsic graded geometric definition of two commuting partial Hodge star operators~$\ast$ and~$\widetilde\ast$, mapping $(p,q)$ tensors to $(D-p,q)$ and $(p,D-q)$ ones respectively \cite{Hull}, and examine the relation of~$\star$ to the product~$\ast\widetilde\ast$. This non-trivial relation is used in several instances in the following. 

Furthermore, drawing a parallel with previous related work in Refs. \cite{Bekaert:2002dt,deMedeirosHull1,Bekaert:2003az,deMedeirosHull2,deMedeiros:2003qel}, we discuss a number of additional geometric operations for bipartite tensors in our formalism, including differentials, traces and their corresponding dual operations with respect to the partial Hodge star operators. Specifically, the dual trace operator allows to define irreducible bipartite tensors under the general linear group, as reviewed in Section \ref{sec23}.
In Section \ref{sec3} we present a unified formula for kinetic and mass terms for arbitrary (Abelian) bipartite tensors in flat spacetime in $D$ dimensions. We show how a number of familiar cases is reproduced, including massless and massive scalars, the Maxwell and Proca actions, the linearized Einstein-Hilbert and Fierz-Pauli actions, and the Curtright action.

For completeness, in Section \ref{sec24} we briefly review the construction of higher derivative (Galileon) interaction terms of general $(p,q)$ bipartite tensors in the formalism of graded geometry. Gauge symmetries are discussed and we prove a higher version of the Poincar\'e lemma (cf. Refs. \cite{DuboisViolette1,DuboisViolette2,Bekaert:2002dt}) in terms of graded geometry, suitable for bipartite tensors, filling a gap in the presentation of Ref. \cite{Chatzistavrakidis1} where the most general symmetries for the higher derivative interaction terms were given without proof. 

In Section \ref{sec4} we turn to the study of standard and exotic dualizations. Our main result is that we find a two-parameter parent Lagrangian that unifies the ones leading to standard duals of differential forms, standard duals of $(p,1)$ bipartite tensors (``generalized gravitons''), exotic duals of differential forms and double duals for generalized gravitons. This Lagrangian is given by \eqref{master} and we repeat it here for completeness:
\be
  \mathcal{L}^{(p,q)}_{\text{P}}(F,\l)=\int_{\theta,\chi}F_{p,q}\star \mathcal{O} \,F_{p,q}+\int_{\theta,\chi}\dd F_{p,q}\ast\widetilde\ast\,\lambda_{p+1,q}\,.\nn
\ee
The notations are explained in detail in the main text; here we note that $F$ and $\l$ are independent reducible bipartite tensors with the indicated degree---equivalently functions of this degree on a graded supermanifold---and the Berezin integration is over two different sets of odd coordinates. The Lagrangian depends on the parameters $p,q\in\N$, such that $p+q\le D-1$, where $D$ is the spacetime dimension. This Lagrangian is universal, in the sense that it contains all possible bipartite duals of a differential form or generalized graviton original field.{\footnote{In the present setting, universality is weak in the sense that it does not include uniqueness.}} For example, when $p+q=3$, the possible partitions lead to the standard and exotic dual of a 2-form and the dual graviton, all from the same general starting point. For $p+q=4$, one obtains the standard and exotic duals of a 3-form and the two different standard duals of the Curtright field. These and more examples are briefly discussed in Section \ref{sec43} in our formalism. Moreover, in Section \ref{sec44} we discuss the extremal case of $p=0$ and its relation to codimension one domain walls. 

The above main result is summarized in Proposition \ref{proposition} and Theorem \ref{theorem}. These establish a common origin of several results known in the literature, and moreover generalize them to the full extent of four domains of values for the parameters $p$ and $q$ and their corresponding dual fields, in particular 
\begin{center}\boxed{
		\begin{tabular}{cccccc}
		Domain  &	{$p$} & {q} & {Original field} & {Dual field} & {Duality type}
			\\[4pt]\hline\\[-4pt]
			I & $\in [1,D-1]$ & $0$ & $[p-1,0]$ & $[D-p-1,0]$ & Standard
			\\[4pt]
			II & $\in [2,D-2]$ & $1$ & $[p-1,1]$ & $[D-p-1,1]$ & Standard
			\\[4pt]
			III & 1 & $\in [1,D-2]$ & $[0,q]$ & $[D-2,q]$ & Exotic
				\\[4pt]
				IV & 2 & $\in [2,D-3]$ & $[1,q]$ & $[D-3,q]$ & Standard
	\end{tabular}}
\end{center}
Here, $[p,q]$ denotes an irreducible $(p,q)$ bipartite field, while a zero entry means that the field is simply a differential form. 

Finally, we include two Appendices. In Appendix \ref{appa} we collect a number of identities and additional proofs. In Appendix \ref{appb} we present the additional necessary details for the completion of the proof of Theorem \eqref{theorem}. 

\section{Mixed symmetry tensors in graded geometry} 
\label{sec2}

\subsection{Differential forms and bipartite mixed symmetry tensors}
\label{sec21}

Let us recall{\footnote{For more details in what follows, the reader may consult Ref. \cite{Qiu:2011qr}.}} that a (smooth) supermanifold ${\cal M}$ is the global object which, for $U$ being an open subset of $\R^{D}$, is locally isomorphic to $C^{\infty}(U)\otimes\bigwedge^{\bullet}(\R^{d})^{\ast}$, 
where the second factor is the exterior algebra. 
${\cal M}$ is thus assigned a dimension $(D,d)$, which means that locally it can be described by a system of standard commuting \emph{even} coordinates $x$ and a set of anticommuting \emph{odd} coordinates, such that 
\be\label{2.1}
\theta^{i}\theta^{j}=-\theta^{j}\theta^{i}~,
\ee
and thus supermanifolds are based on $\Z_2$-graded geometry.  For our purposes, we consider a finer $\Z_d$-grading and focus on $d=D$ and therefore $i=1,\dots,D$. 
This allows to think of differential $p$-forms and $p$-vector fields respectively as degree-$p$ functions over a graded vector bundle, by means of the isomorphisms
\be \label{isos}
C^{\infty}(T[1]M)\simeq \Omega^{\bullet}(M) \quad \text{and} \quad C^{\infty}(T^{\ast}[1]M)\simeq \Gamma(\w^{\bullet}TM)~,
\ee  
where $[1]$ denotes a shift in degree by one and both graded bundles are equipped with degree 0 bosonic coordinates $x^{i}$ and degree~1 fermionic coordinates $\theta^{i}$ and $\chi_{i}$ respectively.
Functions are then expanded accordingly, for example on $T[1]M$
\be 
\o(x,\theta)=\sum_{k=0}^{D}\frac 1{k!}\,\o_{i_1\dots i_k}(x)\,\theta^{i_1}\dots\theta^{i_k}~,
\ee
which is the graded geometry analogue of a collection of $p$-forms with all possible degrees. 

The above correspondence, and the associated integration which may be defined on these graded bundles, is useful for casting physical theories with differential forms into the language of supergeometry. 
However, not all physical theories consist of differential forms only, namely of antisymmetric tensor fields. The most obvious counter example is general relativity in the metric formalism, whose degree of freedom is a symmetric 2-tensor. One then needs to set up the corresponding formalism for multipartite mixed symmetry tensor fields. 
The latter generalize $p$-forms, in the sense that their tensor components have more than one set of antisymmetrized indices.{\footnote{The terminology ``multiforms'' is used in Refs. \cite{Bekaert:2002dt,deMedeirosHull1,Bekaert:2003az,deMedeirosHull2}. We refrain from using this term in this paper.}} 

In the case of bipartite mixed symmetry tensor fields (hence, simply called bipartite tensors) with components having two sets of antisymmetrized indices, one needs  two separate sets of anticommuting variables, say $\theta^{i}$ and ${\chi^{i}}$, which we choose to mutually commute by convention as this turns out to yield simpler expressions,{\footnote{{This choice is purely conventional for terms that are homogeneous in grading. The two species of odd coordinates may be as well chosen mutually anticommuting in exchange for some different signs in various formulas. We shall give an example below, after we introduce some additional structures.}}} 
\be \label{2.6}
\theta^{i}\theta^{j}=-\theta^{j}\theta^{i}~, \quad \chi^{i}\chi^{j}=-\chi^{j}\chi^{i}~,\quad \theta^{i}\chi^{j}=\chi^{j}\theta^{i}~.
\ee   
This means that one considers functions on ${\cal M}=T[1]M\oplus T[1]M$, as in Ref. \cite{Chatzistavrakidis1}. 
The correspondence then is between functions on this graded manifold and bipartite tensors,
\be \label{2.7}
C^{\infty}({\cal M})|_{p,q}=\Omega^{p,q}(M)~, \quad C^{\infty}({\cal M})=\Omega^{\bullet,\bullet}(M)~,
\ee 
where we used the notation 
\be \label{2.8}
\Omega^{\bullet,\bullet}=\bigoplus_{(p,q)\in \Z_+^{2}} \Omega^{p,q}
\ee 
for the corresponding exterior algebra, with $p,q\le D$.. 
This means that an arbitrary $(p,q)$ bipartite tensor may be expanded as{\footnote{ Let us offer three notational hints. In Ref. \cite{Chatzistavrakidis1}, a boldface notation was used for graded quantities to distinguish them from ordinary tensor fields. Since we are not going to use the latter in the present work, we refrain from sticking to this boldface notation. Second, in Ref. \cite{Chatzistavrakidis1} we only worked with covariant mixed symmetry tensors and denoted them as $\o^{(p,q)}$; since we mention contravariant objects in the following, we denote the covariant ones with lower case Greek letters $\o,\z$ with their degree as a (unparenthesized) subscript. 
We  omit the degree of the field and the dependence of the components on the bosonic coordinates whenever they are clear from the context.}} 
\be\label{2.9}
{\o}_{p,q}=\frac{1}{p!q!}\,\o_{i_1\dots i_pj_1\dots j_q}(x)\,\theta^{i_1}\dots\theta^{i_p}\chi^{j_1}\dots\chi^{j_q}~,
\ee
where its tensor components have the mixed index symmetry
\be\label{2.10}
\o_{i_1\dots i_pj_1\dots j_q}=\o_{[i_1\dots i_p][j_1\dots j_q]}~.
\ee 
Evidently, when $q=0$, a $(p,0)$ bipartite tensor is just a $p$-form $\o_p$. Trivial as this might be, it is useful nevertheless, since it is often necessary to view a differential form as a mixed symmetry tensor with a zero slot; we comment further below, when we encounter such situations. 
 
One may then consider a number of operations for bipartite tensors. These operations appear in Refs. \cite{Bekaert:2002dt,deMedeirosHull1,Bekaert:2003az,deMedeirosHull2}, and we will see that they are neatly reformulated in intrinsic graded geometric language. 
The composition of two bipartite tensors is a bilinear map $\O^{p,q}\times \O^{p',q'}\to \O^{p+p',q+q'}$ defined simply by  concatenation
\be\label{2.11}
{\o}_{p,q}\,{\z}_{p',q'}=\frac{1}{p!\,p'!\,q!\,q'!}\,\o_{i_1\dots i_pj_1\dots j_q}\z_{i_{p+1}\dots i_{p+p'}j_{q+1}\dots j_{q+q'}}\,\theta^{i_1}\dots \theta^{i_{p+p'}}\chi^{j_1}\dots \chi^{j_{q+q'}}\,.
\ee
There are two exterior derivatives acting on bipartite tensors, namely ${\dd}:\O^{p,q}\to \O^{p+1,q}$ and $\widetilde{\dd}:\O^{p,q}\to \O^{p,q+1}$ defined as
\be 
\dd=\theta^{i}\partial_i \quad \text{and}\quad  \widetilde{\dd}=\chi^i\partial_{i}~,
\ee 
where $\partial_{i}:=\partial/\partial x^{i}$ is the partial derivative with respect to the bosonic coordinates. 
By construction, they satisfy the identities $\dd^{2}=\widetilde{\dd}^2=0$ and $\dd\,\widetilde{\dd}=\widetilde{\dd}\,\dd$, the latter being controlled by the convention that $\theta$ and $\chi$ commute. (Indeed, had we chosen them anticommuting, we would have obtained that $\dd\,\widetilde{\dd}=-\,\widetilde{\dd}\,\dd$. The latter convention could be used to formulate equivalently everything that follows.)
One can also define an operation which exchanges the two sets of odd coordinates, namely the linear map ${\widetilde{}}\,:\, \O^{p,q}\to \O^{q,p}$ given by
\be\label{tilde}
\o_{p,q}\, \mapsto\,\widetilde{\o}_{q,p}=\frac{1}{p!q!}\o_{i_1\dots i_pj_1\dots j_q}\theta^{j_1}\dots \theta^{j_q}\chi^{i_1}\dots \chi^{i_p}~.
\ee
 This is essentially a transposition map, which in certain cases where confusion may arise we shall denote more explicitly as $\o^{\top_{\theta\chi}}\equiv\widetilde\o$, or when there is no ambiguity also as $\o^{\top}$. Since we are going to write action functionals in flat space, it is worth recalling that we can also cast the Minkowski metric with components $\eta_{ij}=\eta_{(ij)}$ in this formalism as 
\be \label{eta}
\eta=\eta_{ij}\theta^{i}\chi^{j}~.
\ee
Note that $\widetilde{\eta}=\eta$, which is also due to the mutual commutativity of $\theta$ and $\chi$. (The opposite convention would yield $\widetilde{\eta}=-\eta$.)

Moreover, in \cite{Chatzistavrakidis1} we defined a Hodge star operator $\star:\O^{p,q}\to \O^{D-p,D-q}$  for $p+q\le D$ by
\be\label{FullHodge}
	(\star\,\o)_{D-p,D-q}=\frac{1}{(D-p-q)!}\,\eta^{D-p-q}\,\widetilde{\o}_{q,p}\,.\ee 
 The definition is based on the key idea of the graded formalism not to introduce any epsilon tensors ``by hand'' but rather let them arise naturally from the anticommutativity of the graded variables.
	It is observed that this definition contains the tilde operation. As such, even for pure $p$-forms ($q=0$) one needs the second set of anticommuting variables. 
	Thus, this definition is tailored for integration over two sets of odd coordinates, and it is natural for writing down Lagrangians in the graded formalism. 
	
	In addition, one can define two partial Hodge star operations $\ast: \O^{p,q}\to \O^{D-p,q}$ and $\widetilde{\ast}:\O^{p,q}\to \O^{p,D-q}$ in intrinsic graded geometric terms, as follows
	\bea \label{ast1}
	(\ast\,\o)_{D-p,q}&=& \frac 1{(D-p)!}\int\dd^{D}\psi \, \o^{\top_{\theta\psi}}\, (\eta^{\top_{\chi\psi}})^{D-p}~,\\ \label{ast2}
	(\widetilde{\ast}\,\o)_{p,D-q}&=&\frac 1{(D-q)!}\int\dd^{D}\psi \, \o^{\top_{\chi\psi}}\, (\eta^{\top_{\theta\psi}})^{D-q}~,
	\eea 
	where $\omega$ and $\eta$ are the $(p,q)$-tensors in $\theta$ and $\chi$
	as given in equations \eqref{2.9} and \eqref{eta}.
	Let us explain the notation and how the operations work. Here $\psi$ is a third set of auxiliary odd coordinates, aside $\theta$ and $\chi$, which also commutes with the other two sets in our conventions. The Berezin integral is defined in the usual way, 
	\be 
	\int\, \dd^{D}\psi\, \psi^{i_1}\psi^{i_{2}}\dots\psi^{i_D}=\epsilon^{i_1i_2\dots i_D}~,
	\ee 
	for any degree one variable, resulting in the covariant Levi-Civita tensor of  Minkowski space. Note that, in contrast to the standard Levi-Civita symbol, this tensor has the property
	\be\label{e-e}
	\epsilon^{i_1\dots i_p\,k_1\dots k_{D-p}}\epsilon_{i_1\dots i_p\,l_1\dots l_{D-p}}=-p!(D-p)!\delta^{[k_1}_{l_1}\dots\delta^{k_{D-p}]}_{l_{D-p}}\,,
	\ee 
	where the extra minus sign is due to the signature of the Minkowski metric.
	 Moreover, the transposition maps ${\top}_{\th\psi}$ and ${{\top}_{\chi\psi}}$
	 replace a $\theta$ or $\chi$ by a $\psi$ as indicated.
	 To avoid confusion, note that these maps act e.g. on the metric $\eta$ as 
	 \be 
	 \eta^{\top_{\theta\psi}}=\eta_{ij}\psi^i\chi^j\quad \text{and}\quad \eta^{\top_{\chi\psi}}=\eta_{ij}\theta^{i}\psi^{j}\,,
	 \ee
 and similarly for any bipartite tensor. Thus, in \eqref{ast1}, $\omega^{\top_{\theta\psi}}$ has $p$ variables $\psi$ and $q$ variables $\chi$ (and no variables $\theta$), whereas the $(D-p)$-th power of $\eta^{\top_{\chi\psi}}$ has a total of $D-p$ variables $\theta$ and $D-p$ variables $\psi$ (and no variables $\chi$). Therefore their product has $D$ variables $\psi$ (which are integrated out by the Berezin integral), $(D-p)$ $\theta$s and $q$ $\chi$s, thus being a $(D-p,q)$-type bipartite tensor.

	We observe that what has essentially been done is that a bipartite tensor has been treated as a tripartite one with a vanishing extra slot (which we present here in the first position conventionally), namely $(p,q)=(0,p,q)$. Although trivial, this is important for the definition of operations such as the partial Hodge stars above. Then the algorithm for $\ast\,\o_{p,q}$ is: (i) Consider a single auxiliary odd variable $\psi$ that commutes with the variables $\theta$ and $\chi$, (ii) view $\o$ as a tripartite tensor $\o_{0,p,q}$ and transpose it to $\o_{p,0,q}$ with the corresponding operation $\top$, (iii) saturate the $\psi$ variables by means of a suitable transpose of the metric $\eta$  such that the degree of the set that is not to be dualized remains the same, thus obtaining a $(D,D-p,q)$ 
	tripartite tensor, and (iv) integrate over the auxiliary odd variable $\psi$. The result is the partial Hodge star $\ast$. A similar rule holds for $\widetilde{\ast}$. When \eqref{ast1} and \eqref{ast2} are expanded in components in a local coordinate system, the two operations are identical to the ones defined in \cite{Bekaert:2002dt,deMedeirosHull1,Bekaert:2003az,deMedeirosHull2}.
	Two useful properties are the following,
	\be 
	\ast\widetilde{\ast}=\widetilde{\ast}\,\ast \quad \text{and} \quad \ast^2=(-1)^{1+p(D-p)}~,\quad \widetilde{\ast}^2=(-1)^{1+q(D-q)}~,
	\ee 
	acting on any bipartite tensor. 
	
	The operation $\star$ is neither equal nor proportional to the combined action $\ast\,\widetilde{\ast}$. In particular, their relation reads as 
	\be \label{starrelations}
	\star \o=\ast\,\widetilde{\ast}\, \overline{\o}~,
	\ee  
	where 
	\be 
		\overline{\o}:=(-1)^{\e(p,q)}\,\sum_{n=0}^{\text{min}(p,q)}\frac {(-1)^n}{(n!)^2}\,\eta^n\,\text{tr}^n\,\o~,
	\ee 
	where $\e(p,q)=(D-1)(p+q)+pq+1$ for a $(p,q)$ bipartite tensor
	 and $\eta^0 \text{tr}^0 \omega \equiv \omega$. 
It is observed that all traces of $\o$ appear on the right-hand side. Note that the trace is defined in the usual way as 
	\be\label{trace} \text{tr}\,\o_{p,q}=\frac{\eta^{i_1 j_1}}{(p-1)!\,(q-1)!}\,\o_{i_1i_2\dots i_{p}j_1j_2\dots j_{q}}\,\theta^{i_2}\dots\theta^{i_{p}}\chi^{j_{2}}\dots\chi^{j_{q}}\,.
\ee
  We also note in passing that in the special case of $\o$ being a generalized field strength for a bipartite tensor, $\overline{\o}$ is related, but is not identical, to the generalized Einstein tensor defined in Ref. \cite{deMedeirosHull2}. We comment further on this below.
 
Based on the operations $\star$ and $\ast\,\widetilde{\ast}$, we can define two inner products for $(p,q)$-tensors:\footnote{ The definition of the inner product based on $\star$ can be extended to tensors with inhomogeneous degree. It is then invariant under orthogonal transformations of the combined $2D$-dimenstional basis $\{\theta^i, \chi^i\}$ of odd variables. This is a much larger symmetry than the inner product based on $\ast\,\widetilde{\ast}$ enjoys.}
\be \label{biginnerproduct}
\langle \omega_{p,q} , \xi_{p,q} \rangle = \int \, \dd^{D}\theta\,  \dd^{D}\chi\, \omega_{p,q} \star \xi_{p,q} =  
\langle \xi_{p,q}, \omega_{p,q} \rangle \qquad (p+q \leq D)~,
\ee
\be \label{combinedinnerproduct}
( \omega_{p,q} , \xi_{p,q} ) = \int \, \dd^{D}\theta\,  \dd^{D}\chi\, \omega_{p,q} \ast\widetilde{\ast}\, \xi_{p,q} =  
( \xi_{p,q}, \omega_{p,q} ) \qquad (p,q \leq D)~.
\ee
The indicated symmetry of the inner products follows from the symmetry of $\eta$ and the invariance of the integrals under the tilde operation that exchanges $\theta^i$s and $\chi^i$s, see~\eqref{cyclicity}.
Both inner products play an important role in the following: The first inner product allows us to construct geometric Lagrangians  with kinetic terms \eqref{pqkineticlagrangian} that are gauge invariant in the generalized sense discussed in the context of Lemma~\ref{Poincarelemma}.
 The second inner product is instrumental in the construction of the Lagrange multiplier term of the parent Lagrangian \eqref{master} and in the establishment of self-adjointness of the kinetic operator in section~\ref{sec3}.
  
  The partial Hodge star operators may be used to define dual operations to the exterior derivatives and trace. Specifically, the co-differentials are defined as \bea \label{codifferentials}
 && \dd^{\dagger}:=(-1)^{1+D(p+1)} \ast\dd\,\ast: \O^{p,q}\to\O^{p-1,q}~,\\
  &&\widetilde\dd^{\dagger}:=(-1)^{1+D(q+1)} \,\widetilde{\ast}\,\widetilde{\dd}\,\widetilde{\ast}: \O^{p,q}\to\O^{p,q-1}~,
  \eea
  which satisfy the same nilpotency and commutativity relations as the differentials.
  Useful co-trace or dual trace operations may be defined in a similar fashion,
  \bea \label{sigma maps}
  &&\s:=(-1)^{1+D(p+1)}\ast\text{tr}\,\ast: \O^{p,q}\to\O^{p+1,q-1}~,\\
  &&\widetilde\s:=(-1)^{1+D(q+1)}\,\widetilde\ast\,\text{tr}\,\widetilde\ast: \O^{p,q}\to\O^{p-1,q+1}~,
  \eea 
  The dual trace $\sigma$ ($\widetilde\s$) essentially replaces an odd variable of type $\chi$ ($\theta$) with one of type $\theta$ ($\chi$) in the tensor that it acts on. Its expression in components is found to be \cite{deMedeirosHull1}
  \begin{equation*}\begin{split}
  \sigma\,\o=\frac{(-1)^{p+1}}{p!(q-1)!}&\,\o_{\left[{i_1\dots i_{p}}{j_1}\right]j_2\dots j_{q}}\theta^{i_1}\dots\theta^{i_{p}}\theta^{j_1}\chi^{j_2}\dots\chi^{j_q}\,,
  \end{split}
  \end{equation*}
  and similarly for $\widetilde\s$.
   This will assist in the clarification of the irreducibility of the field under $GL(D)$ below. 
   
   Alternatively, one may define the trace, co-traces and co-differentials without reference to the partial Hodge star operators, directly using the graded-geometric variables. To this end, we consider the partial derivatives with respect to the odd coordinates and define 
   \be 
\bar{\theta}_i=\frac{\partial}{\partial\th^{i}}\quad \text{and} \quad \bar{\chi}_i=\frac{\partial}{\partial\chi^i}\,,
\ee
which are assigned degree -1. They behave in the same way as $\theta^i$ and $\chi^i$, namely in our sign conventions 
\be 
\bar{\theta}_{i}\bar{\theta}_j=-\bar{\theta}_j\bar{\theta}_i\,, \quad \bar{\chi}_i\bar{\chi}_j=-\bar{\chi}_j\bar{\chi}_i\,,\quad \bar{\theta}_i\bar{\chi}_j=\bar{\chi}_j\bar{\theta}_i\,.
\ee 
Moreover, their canonical (anti)commutation relations with $\theta^i$ and $\chi^i$ read as{\footnote{We note that had we chosen the opposite sign convention, all relations here would contain anticommutators.}} 
\be 
\{\bar{\theta}_i,\theta^j\}=\{\bar{\chi}_i,\chi^j\}=\d_i^j \quad \text{and}\quad [\bar{\theta}_i,\chi^j]=[\bar{\chi}_i,\theta^j]=0\,.
\ee 
Then the trace and  dual operators defined earlier may be represented as 
\bse \label{alt}
\begin{align}
&\text{tr}=\eta^{ij}\bar{\th}_i\bar{\chi}_j\,, \\[4pt] 
&\s=-\theta^i\bar{\chi}_i\,,\quad \widetilde{\s}=-\chi^i\bar{\th}_i\,,\\[4pt]
& \dd^{\dagger}=\bar{\th}^i\partial_i\,,\quad \widetilde{\dd}^{\dagger}=\bar{\chi}^i\partial_i\,.
\end{align}\ese
   All  operations defined above satisfy several useful identities laid out in Refs. \cite{deMedeirosHull1,Bekaert:2003az,deMedeirosHull2,deMedeiros:2003qel}. We summarize some of them that will be used below in Appendix \ref{appa}, where we also derive and prove several additional ones. 
	
One may alternatively utilize the correspondence of multivectors with functions on the graded bundle $T^{\ast}[1]M$ and define bipartite tensors as functions on 
	$T[1]M\oplus T^{\ast}[1]M$. In that case, the odd coordinates are $\theta^{i}$ and $\chi_{i}$ respectively, and the corresponding bipartite tensors are expanded accordingly as 
	\be 
	{\o}_{p,q}=\frac{1}{p!q!}\,\o_{i_1...i_p}^{j_1...j_q}(x)\,\theta^{i_1}...\theta^{i_p}\chi_{j_1}...\chi_{j_q}~.
	\ee 
	 Then, the metric $\eta$ may be defined simply as 
	 $ 
	 \eta=\theta^{i}\chi_{i}.
	 $
	 We note that the key formulas in the ensuing sections may be written in terms of these variables as well. 
	 
	 Finally, it is worth mentioning that 
	 \be 
	 T[1]M\oplus T^{\ast}[1]M\, \subset \, T^{\ast}[2]T[1]M~, 
	 \ee 
	 the latter containing in addition degree 2 variables $p_i$, the conjugate momenta of the degree~0 coordinates $x^{i}$ with respect to the graded symplectic structure of the graded cotangent bundle appearing above. These momenta and expressions of degree~2 in $\theta^i$ and $\chi_i$ generate degree-preserving symmetries (canonical transformations) via the natural symplectic structure. We shall not pursue this aspect in this paper. 
	 
	 All the above may be generalized in a rather straightforward way to multipartite tensors. Since these are not essential for the main results presented in the rest of the paper, we do not discuss this in detail here.  
	 
\subsection{GL(D) (Ir)Reducibility}
\label{sec23}
So far, with the exception of the metric $\eta$, we have considered bipartite mixed symmetry tensors in $\O^{p,q}(M)$, meaning that their components are only subject to the defining index symmetry (\ref{2.10}) and, therefore transform reducibly under $GL(D)$. However, it is more natural to consider tensors with components corresponding to irreducible representations of $GL(D)$, while, in general, still being reducible under the Lorentz subgroup $SO(D-1,1)$. This can be achieved if one imposes some algebraic constraints on the bipartite tensors, which can also be viewed as imposing an additional index symmetry to their components. 

 Following Refs. \cite{deMedeirosHull1,deMedeirosHull2}, a bipartite tensor is said to be in the $GL(D)$-irreducible subspace
 $\O^{[p,q]}\subseteq\O^{p,q}$ and is represented by $\o_{[p,q]}\in\O^{[p,q]}$ if and only if for $p\ge q$ it satisfies the algebraic constraints
\be\label{2.22}
\s\,\o=0\,,
\ee
likewise for $p\le q$ with $\s$ replaced by $\widetilde \s$ and
\be\label{2.23}
\widetilde\o=\o\quad \text{for}\,\,\,\,\,\,\,p=q\,.
\ee
The subspace $\O^{[p,q]}$ coincides with the whole space $\O^{p,q}$ only for scalars and differential forms.
These constraints constitute the so-called Young symmetry, which is why the $GL(D)$-irreducible subspace $\O^{[p,q]}$ is sometimes termed as the space of $GL(D)$ Young tableau representations with two columns of respective lengths $p$ and $q$.  

Given a general bipartite tensor $\o_{p,q}$, there is a way to obtain a unique irreducible bipartite tensor $\o_{[p,q]}$ by acting with the Young projector $\mathcal{P}_{[p,q]}:\O^{p,q}\to \O^{[p,q]}$ defined by
\bea\label{2.24}
\o_{[p,q]}=\mathcal{P}_{[p,q]}\,\o_{p,q}~.
\eea 
According to Ref. \cite{deMedeiros:2003qel}, this projector takes the form:
\be\label{youngp}
\mathcal{P}_{[p,q]}=\left\{
\begin{array}{ll}
	\mathbb{I}+\sum\limits^{q}_{n=1}c_{n}(p,q)\widetilde \s^n\s^n\,, & p\geq q\\
	\\
	\mathbb{I}+\sum\limits^{p}_{n=1}c_{n}(q,p)\s^n\widetilde \s^n\,, & p\leq q\\
\end{array} 
\right. 
\ee
where 
\be 
c_{n}(p,q)=\frac{(-1)^n}{\prod\limits_{r=1}^nr(p-q+r+1)}~.
\ee The $p=q$ case is covered by both entries due to the first identity in \eqref{so4}
which implies that the two dual trace operators commute for $p=q$. 
The Young projection is by construction idempotent for any bipartite tensor, i.e. $\mathcal{P}_{[p,q]}\o_{[p,q]}=\o_{[p,q]}$.
Evidently, the projection is trivial for $\text{min}(p,q)=0$, since a differential form is already an irreducible representation of $GL(D)$.

\subsection{Kinetic and mass terms}
\label{sec3}

In this section, we briefly review a  subset of the results in Ref. \cite{Chatzistavrakidis1} about kinetic terms for bipartite tensors in graded geometry and extend them to include mass terms.  In the following, we present Lagrangians rather than action functionals. All actions may be obtained by simply integrating over the bosonic coordinates, $\int \dd^{D}x$. 
For simplicity, we  introduce the  shorthand notation 
\be 
\int_{\theta,\chi}:=\int\dd^D\th\,\dd^{D}\chi \nn
\ee
for the integration over odd coordinates.

There is an elegant compact way to write kinetic terms for arbitrary $(p,q)$ bipartite tensor gauge fields using the inner product \eqref{biginnerproduct} that is based on the full Hodge operator~$\star$ defined in~\eqref{FullHodge}:
\be\label{pqkineticlagrangian}
\boxed{{\cal L}_{\text{kin}}(\o_{p,q})=
	\int_{\theta,\chi} \,\dd \omega\,\star\dd\omega\,,}
\ee
 To consider $GL(D)$-irreducible bipartite tensors one has to substitute $\o_{p,q}$ with its irreducible component $\o_{[p,q]}=\mathcal{P}_{[p,q]}\,\o_{p,q}$. This Lagrangian unifies the kinetic terms for any scalar, $p$-form or arbitrary $(p,q)$ bipartite tensor, up to an overall numerical factor. 
Indeed, for the simplest case $p=q=0$, the kinetic term for a scalar field $\o_{0,0}:=\phi$ is obtained,
\be\label{scalarkin}
{\cal L}_{\text{scalar}}(\phi)=-\frac 1{2(D-1)!}\int_{\theta,\chi}\,\eta^{D-1}\,\phi\,\dd\widetilde\dd\,\phi\,,
\ee
where we performed an integration by parts and expressed $\star$ explicitly.
This is the standard scalar kinetic term in the graded formalism, namely ${\cal{L}}_{\text{scalar}}(\phi)=\sfrac 12\phi\Box\phi$, and it indicates that one may define the d'Alembertian as 
\be 
\Box=\partial^i\partial_i:=-\frac 1{(D-1)!}\int_{\theta,\chi}\,\eta^{D-1}\,\dd\widetilde{\dd}\,.
\ee
Similarly,  the kinetic term for an 1-form $\o_{1,0}:=A$ in $D$ dimensions is given by
\be\label{3.2}
{\cal L}_{\text{Maxwell}}(A)=\frac 1 {2(D-2)!}\int_{\theta,\chi} \,\eta^{D-2}\, A \,\dd\widetilde\dd \, \widetilde A\,,
\ee
with $A=A_i(x)\theta^i$, which is equal to the Maxwell Lagrangian in four dimensions, namely $-\sfrac 14 F_{ij}F^{ij}$. 
This kinetic term is obviously gauge invariant under the transformation
$\delta A=\dd\Lambda$
for scalar gauge parameter $\Lambda$. The extension to any Abelian $p$-form gauge field $\o_p$ is  straightforward. The kinetic term will be of exactly the same form as (\ref{3.2}) with $D-p-1$ powers of $\eta$ (and a suitable prefactor) and it will be invariant under the corresponding gauge transformation with respect to a $(p-1)$-form gauge parameter.

In the same spirit, one can consider $p=q=1$, which corresponds to the linearised Einstein-Hilbert action for the metric fluctuation  $h:=h_{[1,1]}=h_{(ij)}\,\theta^i\,\chi^j$. More generally, one may start with a $GL(D)$-reducible $(1,1)$ bipartite tensor $e:=e_{1,1}$, whose tensor components contain both a symmetric part $h$ and an antisymmetric part $b$. The Lagrangian is
\be\label{3.4}
{\cal L}(e)=\frac 1{4(D-3)!}\int_{\theta,\chi} \,\eta^{D-3}\,e\, \dd\widetilde\dd\,\widetilde e\,,
\ee
and it essentially describes the kinetic term for two fields, a graviton and a 2-form, the latter obtained as $-\sfrac 1{12}H_{ijk}H^{ijk}$, where $H=\dd b$.
However, the ordinary graviton corresponds to a $GL(D)$-irreducible bipartite tensor, obtained as $h=\mathcal{P}_{[1,1]}\,e$, which has only symmetric components.{\footnote{The notation $h$ should not be confused with the trace of the graviton, which we always denote as $\text{tr}\,h$ or $h^i{}_{i}$ in components.}} Thus, the familiar component form of the linearized Einstein-Hilbert action, 
\be 
{\cal L}_{\text{LEH}}(h)=-\frac 14\left(h^i{}_i\Box h^j_{\,\,j}-2h^k{}_k\partial_i\partial_j h^{ij}+2h_{ij}\partial^j\partial_kh^{ik}-h_{ij}\Box h^{ij}\right)~,
\ee
 containing only the graviton, would simply be
\be\label{3.5}
{\cal L}_{\text{LEH}}(h)=\frac 1{4(D-3)!}\int_{\theta,\chi} \,\eta^{D-3}\,  h\, \dd\widetilde\dd\, h\,.
\ee
This shows how the formalism based on graded geometry can neatly package several terms including traces and divergences of fields in a single geometric term. This is particularly advantageous when many such terms appear, since they may simply be found upon performing Berezin integration, while making the gauge symmetry of the field completely manifest. Indeed, the gauge symmetry of the Lagrangian (\ref{3.4}) 
\be\label{3.6}
\d e_{1,1}=\dd\k_{[0,1]}+\widetilde\dd\k'_{[1,0]}+c_{i_0i_1i_2}\,x^{i_0}\,\theta^{i_1}\,\chi^{i_2}\,,
\ee 
where $c_{ijk}$ is a constant totally antisymmetric tensor---which will be explained in more generality below,---is also a symmetry of (\ref{3.5}), but the transformed tensor will no longer be manifestly irreducible. In order to manifestly preserve irreducibility we can restrict the transformation to the standard linearized diffeomorphism invariance of  (\ref{3.5}):
\be\label{3.7}
\d h_{[1,1]}=\mathcal{P}\d e_{1,1}=\dd\xi_{[0,1]}+\widetilde\dd\widetilde\xi_{[1,0]}\,.
\ee
We observe that in writing down the graded form of these familiar statements, it is necessary to differentiate between 1-forms defined with respect to $\theta$ and $\chi$. In other words, 1-forms of type $(1,0)$ and $(0,1)$ are both utilized and this is a further reason why it is useful to think of 1-forms (and $p$-forms for that matter) as bipartite tensors with a trivial slot. 

Thus the most interesting and economical feature of the Lagrangian \eqref{pqkineticlagrangian} is that it provides a universal formula for the kinetic term of $(p,q)$ mixed symmetry fields (including scalars and $p$-forms of course) 
with the correct relative factors among different terms that appear in the component expression of the corresponding gauge theories. As a further non-trivial example to test this statement, one can take $(p,q)=(2,1)$ in $D=5$ and expand \eqref{pqkineticlagrangian} in components. 
After some calculation, one finds for the irreducible field 
\bea \label{curtright1}
{\cal L}_{\text{kin}}(\o_{[2,1]})&=& \frac 12 \left(\partial_i\o_{jk|l}\partial^i\o^{jk|l}-2\partial_i\o^{ij|k}\partial^l\o_{lj|k}-\partial_i\o^{jk|i}\partial^l\o_{jk|l}\,-\right. \nn\\ &&\left.-\, 4\o_{i}{}^{j|i}\partial^k\partial^l\o_{kj|l}-2\partial_i\o_{j}{}^{k|j}\partial^i\o^l{}_{k|l}+2\partial_i\o_j{}^{i|j}\partial^k\o^{l}{}_{k|l}\right)~.
\eea 
This is nothing but the Curtright Lagrangian \cite{Curtright:1980yk}, with the correct relative coefficients. Any Lagrangian of this type is contained in \eqref{pqkineticlagrangian}.

Finally, note that in the irreducible case, one may express \eqref{pqkineticlagrangian} in an alternative interesting form. Defining 
\be \label{einstein tensor}
 E_{[p,q]}:=(-1)^{(D-1)(p+q)}\ast\,\widetilde\ast\,\dd\star\dd\,\o_{[p,q]}~,
\ee
which is nothing but the graded geometric expression of the generalized Einstein tensor introduced in \cite{deMedeirosHull1} and in more detail in \cite{Bekaert:2003az,deMedeirosHull2}, one can write
\be\label{pqkineticlagrangian2}
{{\cal L}_{\text{kin}}(\o_{[p,q]})=
	\int_{\theta,\chi} \, \omega_{[p,q]}\,\ast\,\widetilde\ast \,E_{[p,q]}\,.}
\ee
This alternative form allows us to identify the general kinetic operator that acts on bipartite tensor fields: 
 Using $\ast \,\widetilde{\ast}$, the kinetic term \eqref{pqkineticlagrangian} can be rewritten in terms of a degree-preserving kinetic operator $K:\O^{p,q}\to\O^{p,q}$, defined as
\be \label{kinetic operator}
K:=\ast\,\widetilde\ast\,\dd\star\dd\,,
\ee
acting on $(p,q)$ tensor fields. This operator is
 self-adjoint with respect to the inner product $(\omega_{p,q},\xi_{p,q})=\int_{\theta,\chi}\omega_{p,q}\ast\widetilde{\ast}\, \xi_{p,q}$ defined in~\eqref{combinedinnerproduct}. 
 Using the cyclicity property \eqref{cyclicity} and integrating by parts,  
\be \int \omega_{p,q}\,\dd\star\dd \,\xi_{p,q}=\int \xi_{p,q}\,\dd\star\dd\, \omega_{p,q}\ee
holds, up to surface terms that  we assume to be zero for all fields involved. 
Applying identities \eqref{integralidentity2} on both sides of this equation gives 
\be 
\int \omega_{p,q}\,\ast\widetilde{\ast}\,K \xi_{p,q}=\int K \omega_{p,q}\,\ast\widetilde{\ast}\,\xi_{p,q}\,,
\ee 
which is the self-adjointness condition of the kinetic operator $K$.

\color{black}
Introducing mass terms in the same unified spirit is a straightforward task, although it was not presented in Ref. \cite{Chatzistavrakidis1}. 
A general mass term to be added to the general kinetic term (\ref{pqkineticlagrangian}) is
\be\label{3.12}
\boxed{{\cal L}_{\text{mass}}(\o_{p,q})= {m^2}\int_{\theta,\chi}\o\star\o\,.}
\ee
Once again one observes that this single term unifies all known cases and any other case of bipartite tensor.
For example, a free massive scalar field is obtained for $p=q=0$, whereas
introducing a mass term to the Maxwell Lagrangian (\ref{3.2}) leads to the Proca Lagrangian, which in our formalism reads as 
\be\label{3.10}
{\cal L}_{\text{Proca}}(A)=\frac 1{2(D-2)!}\int_{\theta,\chi}\,\eta^{D-2}\left(A \,\dd\widetilde\dd\,\widetilde A+\frac 2{(D-1)}\,\eta \,m^2A \widetilde A\right)\,,
\ee 
with generalization to arbitrary $p$-form fields being obvious.
The linearized Einstein-Hilbert Lagrangian (\ref{3.5}), containing only the symmetric graviton, can also be extended to contain a mass term. The resulting Lagrangian is
\be\label{3.11}
{\cal L}_{\text{FP}}(h)=\frac 1{4(D-3)!}\int_{\theta,\chi}\,\eta^{D-3}\left(h\, \dd\widetilde\dd\, h+\frac 4{(D-2)}\,\eta \,m^2h^2\right)~.
\ee
and it precisely corresponds to the Fierz-Pauli theory for massive gravitons.

\subsection{Higher derivative interaction terms}
\label{sec24}

Apart from kinetic and mass terms, one can also introduce higher derivative interaction terms, with the requirements that they are gauge invariant and they lead to exactly second order field equations. Special care is needed to guarantee that the interaction terms are not total derivatives, in which case they do not influence the dynamics. 
For scalar fields, such interactions were originally classified in \cite{Nicolis:2008in} and they are called Galileons (see Ref. \cite{Deffayet:2013lga} for a review). Galileons for $p$-forms were studied in \cite{Deffayet:2010zh,Deffayet:2017eqq}, and further generalized to mixed symmetry tensor fields of type $(p,q)$ in \cite{Chatzistavrakidis1}, where the geometric significance of these terms was highlighted. 

For the general $(p,q)$ bipartite tensor, the interaction terms as introduced in \cite{Chatzistavrakidis1} are 
\be\label{bipartiteGal}
\boxed{{\cal L}_{\text{Gal}}(\o_{p,q})=\sum_{n=1}^{n_{\text{max}}}\frac 1{(D-k_{n})!}\int_{\th,\chi} \,\eta^{D-k_n}\,\o\,(\dd\widetilde\dd\,\o)^{n-1}(\dd\widetilde\dd\, \widetilde\o)^{n},}
\ee
where $k_n=(p+q+2)n-1$. Note that for $n=1$ this coincides with the standard kinetic term (\ref{pqkineticlagrangian}), so it can be regarded as an interaction term only for $n\geq 2$. 
Turning to the symmetries, Galileon-type actions for bipartite mixed symmetry tensors are invariant under a transformation $\delta \omega_{p,q}$ if $\dd \widetilde\dd \delta \omega_{p,q} = 0$, i.e. if  $\delta \omega_{p,q}$ is ``bi-closed''. 
By the Poincar\'e lemma, on a contractible patch  any ordinary closed form is exact.{\footnote{Certainly this is the case in Minkowski space, which is the relevant case for this paper.}} Here we need to find out how to generalize these notions to mixed symmetry tensors.
We shall prove the following higher version of the Poincar\'e lemma:\footnote{ Another notion of a higher analog of the Poincar\'e lemma has previously been studied in Ref. \cite{Demessie}, but there is no obvious relation to the present work.}

\begin{lemma}\label{Poincarelemma}
	On a contractible patch of a $D$-dimensional manifold, $\dd \widetilde\dd \,\xi_{p,q} = 0$ implies
	\be
	\xi_{p,q} = \dd \kappa_{p-1,q} + \widetilde\dd \kappa_{p,q-1} + c_{i_1 \ldots i_p k_0 k_1\ldots k_q} \theta^{i_1} \ldots \theta^{i_p} x^{k_0} \chi^{k_1} \ldots \chi^{k_q}
	\label{lemma}
	\ee
	where $c_{i_1 \ldots i_p k_0 k_1\ldots k_q}$ is a constant irreducible tensor, which is totally antisymmetric and hence zero for $p+q \geq D$,  the ``gauge parameters'' $\kappa_{p-1,q}$ and $\kappa_{p,q-1}$ are two different mixed symmetry tensors 
	that are labelled only by their degree-indicating subscripts (to avoid cluttered notation) 
	and are zero by convention whenever either subscript is negative.
\end{lemma} 
This appeared without proof in Ref. \cite{Chatzistavrakidis1}; we present its proof in Appendix \ref{appa}. The gauge symmetries of \eqref{bipartiteGal}, including of course the kinetic term, follow directly from Lemma \ref{Poincarelemma}.

In the special case of irreducible tensors of type $[p,p]$, namely when $p=q$, it holds that $\omega=\widetilde\omega$. This allows for an enhanced set of such higher derivative interactions  
\be\label{bipartiteGal0}
\boxed{{\cal L}_{\text{Gal}}(\o_{[p,p]})=\sum_{n=1}^{n_{\text{max}}}\frac 1{(D-k_{n})!}\int_{\th,\chi} \,\eta^{D-k_n}\,\o\,(\dd\widetilde\dd\, \o)^{n},}
\ee
with $k_{n}=(p+1)n+p$ in that case. Scalar fields also fall in this class for $p=q=0$.   To avoid confusion, note that \eqref{bipartiteGal} can be non-trivial only for even appearances of the field $\omega$, in particular $2n$, while \eqref{bipartiteGal0} includes also cases with odd appearances of the field, in general parametrized as $n+1$. This also means that setting $p=q$ in \eqref{bipartiteGal} does not immediately yield all possible terms in \eqref{bipartiteGal0}; it does only under the redefinition $2n \mapsto n+1$.  

Regarding triviality, there are two separate considerations, both turning out to be very simple to check in the graded formalism. First, the maximum value of field appearances in the Lagrangian is fixed against the spacetime dimensionality, since beyond this value the term would vanish identically. In general it is such that the inequality 
$ 
k_{n}\le D $ 
is saturated. For the general case \eqref{bipartiteGal}, this is 
\be \label{nmax}
n_{\text{max}}^{(p,q)}=\left\lfloor \frac {D+1}{p+q+2} \right\rfloor~,
\ee  
while in the particular set of cases \eqref{bipartiteGal0} it becomes{\footnote{Note that this is not 
		equal to \eqref{nmax} with $p=q$  due to the different definition of $n$ in the two cases.}} 
\be \label{nmax0}
n_{\text{max}}^{[p,p]}=\left\lfloor \frac {D-p}{p+1}\right\rfloor~.
\ee 
Second, it is easy to prove that \cite{Chatzistavrakidis1}
\be 
(\dd\widetilde\dd\, \o_{p,q})^2|_{p+q=\text{odd}}=0=(\dd\widetilde\dd\, \widetilde\o_{q,p})^2|_{p+q=\text{odd}}~.
\ee 
This means that for the total degree $p+q$ being odd all such interactions are total derivatives, thus they do not contribute to the field equations and may be ignored. 
This is the case for instance for a Maxwell field, which is a $1$-form, or for the Curtright field, which can be represented by a $GL(D)$-irreducible bipartite tensor $[2,1]$. However, such interactions do exist for the graviton, and they are the familiar Lovelock terms in any dimension, as explained in detail in Ref. \cite{Chatzistavrakidis1}. (For the graviton in particular, the relevant Lagrangian is the enhanced one in Eq. \eqref{bipartiteGal0}.)
Also, these interactions exist for a 2-form, for example for the Kalb-Ramond field of string theory, which is expected since at higher orders in the $\a'$-expansion one finds higher derivative interactions. Moreover, more possibilities may arise for mixed interactions between different species of tensor fields, in which case even the previously forbidden fields may participate. More details are found in \cite{Chatzistavrakidis1}.

\section{Standard and exotic dualization} 
\label{sec4}

\subsection{Generalities and terminology}
\label{sec41}

The tensor fields discussed in Section \ref{sec2}  correspond to particular representations of the general linear group $GL(D)$ and its Lorentz subgroup $SO(D-1,1)$, which are in general reducible. 
If we  demand in addition that they have Young symmetry, then they  correspond to \emph{irreducible} representations of $GL(D)$. However, in any covariant gauge theory involving such gauge fields, the physical field with the true propagating degrees of freedom is found after gauge fixing. The field equations for the physical field in the free theory  reduce to the free wave equations $\Box\,\o=0$ and no gauge invariance remains in the theory. 

In the physical theory, the field should not only be $SO(D-1,1)$-irreducible but also irreducible under the little group $SO(D-2)\subset SO(D-1,1)\subset GL(D)$. This basically means that the physical field will have to be fully traceless with respect to the $SO(D-2)$-invariant metric. In addition, it will be represented by a tensor with the same number of indices (and index symmetries) as in the pre-gauge-fixed case, but now these indices will run from $1$ to $D-2$. These being said, an arbitrary physical theory corresponding to a particular irreducible tensor representation of $SO(D-2)$ can arise from a number,  infinite according to Ref. \cite{Boulanger:2015mka}, of different covariant gauge theories. 
The key feature here is that different irreducible representations of $GL(D)$ may correspond to equivalent irreducible representations of $SO(D-2)$, after full gauge fixing.

In the simplest setting, one considers a gauge theory for a $p$-form and another gauge theory for a $(D-p-2)$-form, which are both irreducible representations of $GL(D)$. These gauge fields have different number of components, namely $\binom{D}{p}$ and $\binom{D}{D-p-2}$. 
However, after gauge fixing, the physical field in the first case is a $p$-form representation of $SO(D-2)$ with $\binom{D-2}{p}$ components, while in the second case it is a $(D-p-2)$-form representation of $SO(D-2)$ with $\binom{D-2}{D-p-2}$ components. Therefore the degrees of freedom of the two physical fields end up being the same, which was to be expected since they are related by Hodge duality. 
Thus, the initial covariant gauge theories are physically equivalent, or electric/magnetic duals of one another.

As discussed in the introduction, the dynamics of dual fields may be treated on an equal footing by means of a parent (or master) action. This action contains two fields, one of which is a Lagrange multiplier; when it is integrated out, the original theory is obtained. Alternatively, when one integrates out the other field, a dual theory is obtained. Although the standard case described in the previous paragraph is well-known, in recent years it was realized that a field does not necessarily have only a single dual. Indeed, $p$-forms have exotic duals too, bipartite tensors have not only exotic duals but also double duals \cite{Hull,deMedeirosHull1,Boulanger1,Boulanger2}, while in general there is an infinite number of ways to represent a physical field \cite{Boulanger:2015mka}. 

 In this generalized sense, duality is not always realized at the level of Lagrangians; it happens that the dual Lagrangian often contains additional fields aside the dual gauge field. 
 One then should find the correct dual field equations that correspond to the same propagating physical degrees of freedom. 
 In the following we employ this general perspective and consider two gauge theories dual when they can be obtained by the same parent action on-shell and they imply  field equations for two different tensor fields in  irreducible representations of $GL(D)$ that are equal in number after complete gauge fixing.

Starting with an original $N$-partite tensor, namely one of type $[p_1,\dots,p_{N}]$, we use the following terminology:
\begin{enumerate}
		\item \textit{$n$-Standard dual}: The $N$-partite tensor is dualized in $n$ selected tensor slots, yielding an $N$-partite tensor of type $[p_1,\dots,D-p_{\a_1}-2,\dots,D-p_{\a_2}-2,\dots,D-p_{\a_{n}}-2,\dots,p_{N}]$. Evidently, a 1-standard dual is just a standard dual, the Hodge dual in the case of differential forms. 
		As an example, the standard dual of the graviton $[1,1]$ is a $[D-3,1]$ bipartite tensor, while its 2-standard (or double) dual is a $[D-3,D-3]$ bipartite tensor. 
	\item \textit{$n$-Exotic dual}: An $(N+n)$-partite tensor of type $[D-2,\dots,D-2,p_1,\dots,p_{N}]$, with $n$ entries of $D-2$. We call a 1-exotic dual simply exotic dual. Exotic duality increases the order $N$ of the tensor field. 
\end{enumerate}  
Note that multiple standard duals may exist, and even multiple $n$-standard duals. For instance, the Curtright field $[2,1]$ has two standard duals $[D-4,1]$ and $[2,D-3]$ respectively, but only one double dual $[D-4,D-3]$. Regarding scalar fields, they have a standard dual only in two dimensions when they are considered as the $N=0$ case; in any other dimension, any dual of a scalar is exotic, for instance its $(D-2)$-form dual is the 1-exotic. When scalars are considered as the $N=1$ case with $p_1=0$, then the $(D-2)$-form dual is the standard dual, as for any $p$-form. This is of course not essential, being just a peculiarity of the terminology for the special case of scalar fields.
Furthermore, the possibility of mixed standard/exotic duals is also available, but we do not introduce separate terminology for such cases.

In the rest of the paper we will deal with 1- and 2-standard duals and 1-exotic duals and show how they may be unified in a single parent Lagrangian. We leave a complete discussion of $(n>2)$-standard and $(n>1)$-exotic duals for a future publication \cite{prep}.

\subsection{A universal parent Lagrangian for bipartite tensors}
\label{sec42}

In the present section, our aim is to find a universal{\footnote{\label{foot} Universality typically refers to and includes uniqueness. In the present context this is neither true, nor desirable. Parent first order actions are known not to be unique. Our use of the word refers here to a parent action encompassing all cases of interest in a certain domain to be defined below. Strictly speaking this is a \emph{weak} universality, although we are not going to use this terminology.}} parent Lagrangian that simultaneously accounts for the following: 
\begin{itemize} 
	\item The standard and exotic duals for any differential $p$-form, and
	\item the standard and double standard duals for any bipartite tensor of type $(p,1)$.  
	\end{itemize} 
In other words, we are looking for a Lagrangian that would be the parent action for any dualization of  differential forms and $(p,1)$ bipartite tensors that does not lead to a $N$-partite field with $N$ greater or equal to three.

Clearly the sought-for Lagrangian must have a parametric dependence, thus being  a family of Lagrangians. We present directly the final result, then prove that it is a correct Lagrangian for our purposes, and  subsequently we test it in a number of relevant and non-trivial cases. 
The $2$-parameter parent Lagrangian we propose reads in $D\geq p+q+1$ dimensions as
\be\label{master}
 \boxed{  \mathcal{L}^{(p,q)}_{\text{P}}(F,\l)=\int_{\theta,\chi}F_{p,q}\star \mathcal{O} \,F_{p,q}+\int_{\theta,\chi}\dd F_{p,q}\ast\widetilde\ast\,\lambda_{p+1,q}\,,}
\ee
where $F_{p,q}$ and $\lambda_{p+1,q}$ are general, independent $GL(D)$-reducible{\footnote{{The reason we work with reducible fields may be explained as follows. In linearised gravity, the gravitational field transforms under local Lorentz transformations according to $\d_{\Lambda}e_{ij}=\Lambda_{ij}$, where $\Lambda_{ij}=-\Lambda_{ji}$ is the (antisymmetric) Lorentz parameter. This is only possible if $e$ is a reducible field, even though eventually only its irreducible component will appear in the second order Lagrangian---its antisymmetric part entering only through total derivative terms. In addition, the field equation for the field $F$ in that case, which will imply the linearised Einstein equation on shell, has Lorentz symmetry only if the fully antisymmetric part of the field $\l$ transforms under Lorentz transformations. However, this fully antisymmetric part is available only when the field $\l$ is chosen reducible. This is first mentioned in \cite{WestL1} and further clarified in \cite{WestL2}. Therefore we always consider reducible fields $F$ and $\l$, unless of course $q=0$, in which case they are both standard differential forms.}} bipartite tensors, $\mc{O}:={\cal O}^{(p,q)}$ is some $p-$ and $q-$dependent operator acting on $(p,q)$ bipartite tensors to be defined below and the subscript $\text{P}$ stands for ``Parent''. Observe that the first term contains the Hodge operator $\star$, while the second term the combined action of the two partial Hodge operators $\ast$ and $\widetilde\ast$. As discussed around Eq. \eqref{starrelations}, the two are not equal but instead they differ by all possible traces of the field.{\footnote{This is of course irrelevant for differential forms, which have no traces anyway.}} Up to an overall factor and arbitrary rescaling of the Lagrange multiplier $\lambda_{p+1,q}$, we will now show that this Lagrangian unifies the known parent Lagrangians in the literature for standard and exotic dualizations and, furthermore, contains parent Lagrangians for the dualization of bipartite gauge fields, such as the Curtright field.

Let us explain the role of the operator $\mc O$ that appears in \eqref{master}. Recovering the original second order action for a given field $\o$ requires variation of \eqref{master} with respect to the Lagrange multiplier $\lambda_{p+1,q}$. The equation of motion that one obtains is the Bianchi identity on $F_{p,q}$, meaning that on-shell
\be \label{Bianchi}
\dd F_{p,q}=0 \, \Rightarrow \, F_{p,q}=\dd\omega_{p-1,q}~,
\ee 
the latter being true locally for some field $\omega_{p-1,q}$. Electric/magnetic duality relates two inequivalent irreducible representations of $GL(D)$; i.e. substituting \eqref{Bianchi} back into \eqref{master} should give a covariant gauge theory containing an irreducible field.
However, unless $q=0$, $\omega_{p-1,q}$ is a reducible field and this is exactly the problem that the operator $\mc O$ solves. 
Indeed, ${\cal O}$ can be determined by the requirement that it must satisfy locally
\bea \label{calorequirement}
\mc O \,\dd\o_{p-1,q}\overset{!}=\dd\o_{[p-1,q]}+\widetilde\dd X~,
\eea
where $X$ is some $(p,q-1)$ field that is acted upon by $\widetilde\dd$. This would then select only the irreducible component of $\o_{p-1,q}$ in \eqref{master}, while collecting at the same time all other components under $\widetilde\dd$. However, bearing in mind that $\dd^2=0$, and that $\star$ contains a tilde operation acting on 
$\mc O F$, the $\widetilde\dd X$ sector corresponds to a total derivative term in \eqref{master} and may be dropped. Thus eventually the Lagrangian will only depend on 
the irreducible component $\o_{[p-1,q]}$ on-shell, as will be proven below. 
	
Following the above logic, denoting $\omega:=\omega_{p-1,q}$ and $[\omega]:=\o_{[p-1,q]}$ and for $p\ge q+1$, we directly compute{\footnote{Eventually only the form of the operator for $q<2$ or $p<3$ will turn out to be relevant for our purposes, however the computation may be done in general.}} 
\bea 
\mc O \,\dd\o-\dd[\o]&=&\mc O \,\dd\o-\dd {\cal P}_{[p-1,q]}\o=\mc O \,\dd\o-\dd\left(\mathbb{I}+\sum^{q}_{n=1}c_{n}(p-1,q)\widetilde \s^n\s^n\right)\o \nn\\
&=&\left(\mc O-\mathbb{I}\right)\dd\o-\sum^{q}_{n=1}c_{n}(p-1,q)\dd\widetilde \s^n\s^n\,\o\nn\\ 
&=&\left(\mc O-\mathbb{I}-\sum^{q}_{n=1}c_{n}(p-1,q)\widetilde \s^n\s^n\right)\dd\o+\widetilde\dd Y~,
\eea
for some calculable $Y$. In the last step, we used the identities \eqref{id2} and \eqref{id3},
and the corresponding transpose ones. 
By \eqref{calorequirement} we conclude that 
\be 
\label{calo1}
{\cal O}=\mathbb{I}+\sum^{q}_{n=1}c_{n}(p-1,q)\widetilde \s^n\s^n~,\quad \text{for}\quad p\ge q+1~.
\ee 
We observe that in the $p\ge q+1$ case $\mc O^{(p,q)}$ is formally the same as $\mc P_{[p-1,q]}$; however, in \eqref{master} it acts on a $(p,q)$ bipartite tensor rather than a $(p-1,q)$, therefore it is not the projector.

The computation proceeds differently in the $p< q+1$ case. A useful identity, which we prove in  Appendix \ref{appa}, is 
\be \label{app1}
\dd\left(\s^n\,\widetilde \s^n+n^2\,\s^{n-1}\,\widetilde \s^{n-1}\right)\o=\s^n \,\widetilde \s^n\dd\o+\widetilde\dd Y~,
\ee 
where $Y$ is once more calculable but irrelevant for our purposes. This formula is essentially a recursion relation, which can be used to prove the intermediate result 
\be \label{app2}
\dd\s^n\,\widetilde \s^n\, \o_{p-1,q}=\s^n\,\widetilde \s^n\,\dd\o_{p-1,q}+\sum_{k=1}^{n}(-1)^k\prod_{m=0}^{k-1}(n-m)^2\s^{n-k}\,\widetilde \s^{n-k}\,\dd\o_{p-1,q}+\widetilde\dd Y'~,
\ee 
as shown in Appendix \ref{appa}. Finally, this can be used to compute ${\cal O}$ via \eqref{calorequirement} and, collecting the previous result \eqref{calo1} too, find
\bea \label{calo}
{\cal O}=\left\{
\begin{array}{ll}
\mathbb{I}+\sum\limits^{q}_{n=1}c_{n}(p-1,q)\,\widetilde \s^n\,\s^n	\,, & p\geq q+1\\
	\\
\mathbb{I}+\sum\limits_{n=1}^{p-1}c_{n}(q,p-1)\left(\s^n\,\widetilde \s^n+\sum\limits_{k=1}^{n}(-1)^k\prod\limits_{m=0}^{k-1}(n-m)^2\s^{n-k}\,\widetilde \s^{n-k}\right)\,, & p< q+1\\
\end{array}~.
\right.
\eea
Note that for $p<q+1$ the result is  more complicated. The reason is that due to the action of the differential $\dd$ on $F_{p,q}$ in the second term of \eqref{master}, $p$ and $q$ are treated asymmetrically. This is necessary in order to account for  all types of duals we are after.

At this stage, solving the field equation for the Lagrange multiplier and using what we have proven so far leads to the on-shell Lagrangian 
\be 
{\cal L}^{(p,q)}_{\l\text{-on-shell}}=\int_{\theta,\chi}\dd\o\star \dd[\o]~,
\ee 
and we notice that the second occurrence of the field contains only its desired irreducible component, however the first does not. Had we attempted to act with ${\cal O}$ also on the first $F_{p,q}$ in \eqref{master} would not solve the problem but rather worsen it, since then we would not have been able to drop the $\widetilde\dd X$ terms using total derivatives. Fortunately, this is unnecessary for the cases mentioned in the beginning of the present section. 
To show this, we calculate for $p\ge q+1$, 
\bea 
{\cal L}^{(p,q)\,;\,p\ge q+1}_{\l\text{-on-shell}}&=&\int_{\theta,\chi}\dd\o\star \dd[\o]
\overset{\eqref{youngp}}=\int_{\theta,\chi}\dd\left([\o]-
\sum\limits_{n=1}^q c_{n}(p-1,q)\,\widetilde \s^n\,\s^n\,\o\right)\star \dd[\o]= \nn\\
&=&\int_{\theta,\chi}\dd[\o]\star \dd[\o]
-{\cal L}_{\text{rest}}~,\label{parentonshell0}
\eea 
where we defined
\be 
{\cal L}_{\text{rest}}=\int_{\theta,\chi}\sum\limits_{n=1}^q c_{n}(p-1,q)\,\dd\,\widetilde \s^n\,\s^n\,\o\star \dd[\o]~.
\ee
Since the first term in the right hand side of \eqref{parentonshell0} depends only on the irreducible component of the field $\o$, it remains to be shown that ${\cal L}_{\text{rest}}$ is zero up to a total derivative. This is not true in general. 
Obviously, it is true for $q=0$ and any $p\in (0,D)$. For $q=1$ and $p\in (1,D-1)$, defining 
$c(p):=\frac{c_1(p-1,1)}{c_1(p,1)}$,
we find 
\bea 
{\cal L}_{\text{rest}}&=&\int_{\theta,\chi}c_1(p-1,1)\, \dd \widetilde \s\,\s\, \o \star \dd[\o]\overset{\eqref{id2}}=\int_{\theta,\chi}c_1(p-1,1)\, \widetilde \s\,\s\,\dd  \o \star \dd[\o] =\nn\\[3pt] &=& c(p)\int_{\theta,\chi}c_1(p,1)\, \widetilde \s\,\s\,\dd  \o \star \dd[\o]=
c(p)\int_{\theta,\chi}\left({\cal P}_{[p,1]}-\mathbb{I}\right)\dd  \o \star \dd[\o]=\nn\\[3pt] 
&=&c(p)\int_{\theta,\chi}\left([\dd  \o]-\dd\o\right) \star \dd[\o]=c(p)\int_{\theta,\chi}\dd\o\, \star \left([\dd[\o]]-\dd[\o]\right)=0~,
\eea
where in the last step we used the following two general relations 
\bea 
\int_{\theta,\chi}[\o]\star \xi=\int_{\theta,\chi}\o\star[\xi] \quad \text{and}\quad [\dd[\o]]=\dd[\o]~. 
\eea
The first relation is proven in \cite{deMedeirosHull2}, while the second one is obvious, since the exterior derivative on an irreducible field yields an  irreducible field too. Thus we have proven that 
\be 
{\cal L}^{(p,q)\,;\,p\ge q+1}_{\l\text{-on-shell}}={\cal L}_{\text{original}}([\o])~, \quad \text{for} \quad q=0,1,
\ee 
where the latter is the  second order Lagrangian for the original irreducible field. 
On the contrary, for $p\ge q+1$ and $q\ge 2$, and without further restrictions, ${\cal L}_{\text{rest}}\ne 0$. The case $p\le q+1$ is completely analogous, and one finds that the corresponding ${\cal L}_{\text{rest}}$ vanishes for $p=1$ and $p=2$.  
Collecting our results, we have proven the following 
\begin{prop} \label{proposition}
	The parent first-order Lagrangian \eqref{master} for $GL(D)$-reducible independent bipartite tensors $F_{p,q}$ and $\l_{p+1,q}$ and with ${\cal O}$ given by \eqref{calo}  is equivalent on-shell with the second order Lagrangian \eqref{pqkineticlagrangian} for a $GL(D)$-irreducible bipartite field $\o_{[p-1,q]}$, for parameters $p$ and $q$ in the domains $\{p\in[1,D-1],q=0\}\cup\{p\in [2,D-2],q=1\}\cup\{p=1,q\in [1,D-2]\}\cup\{p=2,q\in [2,D-3]\}$.
	\end{prop}

We will see that the four domains correspond respectively to the standard dualization of a $p$-form, the standard dualization of a $(p,1)$ generalized graviton, the exotic dualization of a $q$-form and the remaining standard and double dualization of a generalized graviton. Moreover, the extremal $p=0$ case is also admissible, however we will comment on it separately at the end of this section.
	On the other hand, outside the domain where Proposition \ref{proposition} holds, at face value one would obtain a second order action that depends on more fields than just the irreducible components of $\o$. 
	An extension of \eqref{master} or another parent action that would include dualities for fields that do not belong in the above domains is beyond the scope of the present paper. 

We now focus on the procedure of determining the Lagrangian and field equations for the dual theory. 
Let us first consider the domain $\{p\in [1,D-1], q=0\}$, since this is the simplest case. 
Here the parent Lagrangian reads as
\be\label{parentdomain1}
\mathcal{L}_{\text{P}}=(-1)^{\e+1}\int_{\th}F_{p,0}\ast F_{p,0}-\int_{\th}\dd F_{p,0}\ast\l_{p+1,0}\,,
\ee
where $\e$ denotes the parity relating the Hodge $\star$ with the product of partial Hodge stars $\ast\widetilde\ast$,  known from our result in Section \ref{sec2}. Note that in \eqref{parentdomain1} we already performed the  Berezin integration over the second set of odd coordinates, which is trivial in the present case. Variation w.r.t. $F$  gives directly the duality relation
\be\label{dr1} F=\frac{(-1)^{\e+Dp}}{2}\ast\dd\widehat\o\,,
\ee
where we have defined the dual field $\widehat\o_{D-p-1,0}\equiv \ast\l_{p+1,0}$. This is indeed the expected duality relation between the field strengths $F=\dd\omega$ and $\widehat{F}=\dd\widehat{\omega}$ of the two dual differential forms $\omega$ and $\widehat{\omega}$.  Substituting back into $\mathcal{L}_{\text{P}}$, one gets the dual second order Lagrangian
\be
\mathcal{L}_{\text{dual}}(\widehat\o)=\frac 14\int_{\theta,\chi}\dd\widehat{\o}\star\dd\widehat{\o}\,,
\ee 
which is just the kinetic term for the dual field $\widehat{\o}$, the latter being trivially an irreducible field since it is just a differential form. We conclude that in this first case the parent action relates two dual second order ones for differential forms $\o_{p-1}$ and $\widehat{\o}_{D-p-1}$ with the same degrees of freedom under the little group, as already explained.  This is just standard electric/magnetic duality with the exchange of field equations and Bianchi identities for the dual fields. Specifically, by means of the duality relation \eqref{dr1}, the Bianchi identity $\dd F=0$ becomes the equation of motion $\dd^{\dagger}\widehat F=0$, whereas the equation of motion $\dd^{\dagger}F=0$ becomes the Bianchi identity $\dd\widehat F=0$.

For the remaining three domains of $p$ and $q$ values, the situation is more complicated because the dual field is no longer a differential form. We sketch the main steps of the analysis here and leave further details for Appendix \ref{appb}. For the domain $\{p\in [2,D-2],q=1\}$, upon integrating out $F$, we obtain a second order Lagrangian in terms of the field $\l$, which reads as 
\be\label{-dual1}
\mathcal{L}_{\text{dual}}[\l]=\frac{(-1)^{\e+1}}{4}\int_{\theta,\chi}\left(\mathbb{I}-\widetilde \s\s-\frac{1}{D-p}\eta\,\text{tr}\right)\dd^{\dagger}\l\ast\widetilde\ast\,\dd^{\dagger}\l\,.
\ee 
This is not the final dual Lagrangian, since it depends on the reducible field $\l$, which is not the dual field. We proceed by decomposing the Lagrange multiplier as 
\be \label{decomp1}
\l_{p+1,1}=\widehat\l_{p+1,1}+\eta\, \mathring{\l}_{p,0}~, 
\ee 
 where $\widehat{\l}$ is traceless, $\text{tr}\,\widehat\l= 0$. The Hodge dual of any traceless bipartite tensor is by definition $GL(D)$-irreducible, so we can now define the irreducible dual field $\widehat\o_{[D-p-1,1]}$ as $\widehat\l:=\ast\,\widehat\o$. 
 
The duality relation obtained upon variation w.r.t. $F$ leads on shell to (see Appendix \ref{appb} for the derivation) 
 \be \label{dr2}
R=\frac{(-1)^{D(p+1)}}{2}\ast \widehat{R}\,,
\ee 
where we defined the irreducible (generalized) Riemann tensors 
\be 
R:=R_{[p,2]}=\dd\widetilde{\dd}\,\omega_{[p-1,1]} \quad \text{and}\quad \widehat{R}:=\widehat{R}_{[D-p,2]}=\dd\widetilde{\dd}\,\widehat{\omega}_{[D-p-1,1]}\,.
\ee 
We observe that the duality relation obtained from the Lagrangian is of the desired form, relating the field strengths of the two dual (irreducible) fields by partial Hodge duality. The Riemann tensors satisfy Bianchi identities $\dd R=0=\widetilde{\dd}R$, equations of motion of Einstein type $\text{tr}R=0$ and the irreducibility criterion $\s R=0$, respectively for $\widehat{R}$. It is instructive to note how Bianchi identities and equations of motion are mapped into each other under duality in this case, since it is slightly different than the simple case of differential forms. The duality relation \eqref{dr2} shows that the Bianchi identity $\widetilde{\dd}R=0$ implies the Bianchi identity $\widetilde{\dd}\widehat{R}=0$ simply because the operator $\widetilde{\dd}$ commutes with the operator $\ast$. On the other hand, the irreducibility of $R$, $\s R$=0, is mapped under duality to $\s\ast \widehat R=0 \Leftrightarrow \text{tr}\widehat{R}=0$, by definition of $\s$. In other words, irreducibility of $R$ implies the Einstein equation for $\widehat R$ and vice versa. Finally, the Bianchi identity $\dd R=0$ and the equation of motion $\text{tr}\,R=0$ imply the Bianchi identity $\dd\widehat{R}=0$.

Returning to the off shell parent Lagrangian, using the decomposition \eqref{decomp1} of $\l$, we show in full generality in Appendix \ref{appb} that the $\mathring{\l}$-dependence in the dual Lagrangian \eqref{-dual1} drops out algebraically and one gets again the standard kinetic term for the dual field $\widehat\o_{[D-p-1,1]}$. 
 Thus the two theories for $\o_{[p-1,1]}$ and $\widehat\o_{[D-p-1,1]}$ are obtained on-shell from the parent Lagrangian \eqref{master} for this domain; moreover, the two fields have the same number of components under $SO(D-2)$ and therefore they obey the same number of field equations after complete gauge fixing, as required. 

In the case when the domain of values is $\{p=1,q\in [1,D-2]\}$, a different complication arises. 
The dual Lagrangian in terms of $\l$ is found to be 
\be\label{dual3}
\mathcal{L}_{\text{dual}}[\l]=\frac{3(-1)^{\e}}{8}\int_{\theta,\chi}\left(\mathbb{I}-\frac{1}{D-q}\,\eta\,\text{tr}\right)\dd^{\dagger}\l\ast\widetilde\ast\, \dd^{\dagger}\l\,.
\ee 
As in the previous case, it depends on a reducible field $\l_{2,q}$, which we decompose as \be\label{dec} 
	\lambda_{2,q}=\widehat\l_{2,q}+\eta \mathring{\l}_{1,q-1}~,
	\ee
	where $\text{tr}\widehat\l= 0$. Then one defines the $GL(D)$-irreducible dual field $\widehat\o_{[D-2,q]}$ via the relation $\widehat \l=\ast\,\widehat \o$. As discussed in more detail in Appendix \ref{appb}, the on shell duality relation obtained from the Lagrangian in this case leads to 
\be 
\label{dr3} R=\frac{(-1)^{qD}}{2}\ast \widehat{R}\,,
\ee where the Riemann tensors are defined as before, albeit for the fields relevant for the present case---therefore they have different degree than the second domain, presently being $R=R_{[1,q+1]}$ and $\widehat{R}=\widehat{R}_{[D-1,q+1]}$. Nevertheless, we observe that once more we obtain an equation that relates the  field strengths of the two irreducible dual fields by partial Hodge duality. However, unlike the previous two cases, the dual fields are not of the same type, the first being a differential form and the second a bipartite tensor, thus they do not obey the same type of equations of motion. Indeed, the original field obeys the Bianchi identities and Einstein equation $\text{tr}\,R=0$ as before,{\footnote{It is useful to note here that this ``Einstein'' equation can be easily translated in the more familiar form of the field equation for a differential form as follows. Using \eqref{id7}, it is obviously equivalent to $\widetilde{\dd}^{\dagger}G=0$, where $G=\widetilde{\dd}\omega$ is the standard ``Maxwell'' form of the field strength of $\omega$.}} and the irreducibility condition $\widetilde{\s}R=0$, since for this domain $p\le q$. Unlike the second domain, the irreducibility condition implies that $\text{tr}\,\widehat{R}\propto \ast\,\dd\widetilde{\dd}\,\s\,\omega\ne 0$, therefore the passage from the equations that $R$ satisfies to the ones that $\widehat{R}$ satisfies using the duality relation is more subtle. Indeed, relation \eqref{dr3} leads to an equation of motion for the dual field only upon taking higher traces of the generalized Riemann tensor, in particular 
\be \label{trq}
\text{tr}^{q+1}\widehat{R}=0\,.
\ee
These are field equations for a single irreducible field and they are equal in number with the ones for the original field $\o_{[0,q]}$ after complete gauge fixing.
However, they are not obtained at face value from the equations of the original field through the duality relation, unlike the standard cases laid out before. In addition, they cannot be obtained as such from a Lagrangian as Euler-Lagrange equations. Thus the only way that this duality can be realized off shell, in particular that the duality relation \eqref{dr3} follows from the Lagrangian, is by means of additional fields that are not dynamical on shell but cannot be algebraically eliminated from the Lagrangian, as discussed for special cases in \cite{Boulanger:2015mka,Bergshoeff}. This is a feature that distinguishes this case from the previous two. 

The above feature reflects itself in the Lagrangian \eqref{dual3} as follows. Substituting the decomposition \eqref{dec} in it, the trace field $\mathring{\l}$ does not cancel out algebraically this time.
The same is true for the equations of motion obtained from it, which may be written as
\be\label{eoms1234}
\text{tr}\,\widehat{R}=(-1)^{D+1}\ast\dd\widetilde{\dd}\left(\mathbb{I}-\frac 1{D-q}\eta\,\text{tr}\right)\mathring\l-\widetilde{\dd}\,\text{tr}\,\dd\,\widehat\omega\,,
\ee  
and they contain both the dual field $\widehat\o$ and the trace field $\mathring{\l}$.  It is indeed observed that $\text{tr}\,\widehat{R}\ne 0$, as expected from the earlier discussion. This is then the exotic theory connected to the $q$-form theory through the parent Lagrangian  \eqref{master} which implements the exotic duality relation \eqref{dr3}.  
The mechanism to eliminate the additional off shell fields on shell was suggested in \cite{Boulanger:2015mka} and in the present case it simply amounts to acting on both sides of \eqref{eoms1234} with the operator $\text{tr}^q$. Then one is left with an 
equation of motion just for the dual field $\widehat\o$, i.e.
\be\label{eomcorrect}
 \text{tr}^{q+1}\,\dd\widetilde\dd\,\widehat\o_{[D-2,q]}=0\,,
\ee
 which is identical to \eqref{trq}. We observe that there is a difference between the on shell and the off shell duality statements; on shell, the original and dual theories have the same spectrum of fields as explained below \eqref{dr3}, however the off shell exotic theory contains additional fields, as already noted in \cite{Boulanger1, Boulanger2}. 
 Similar statements hold in the final case of the domain $\{p=2,q\in [2,D-3]\}$, discussed in detail in Appendix \ref{appb}. 
 
 Collecting all the above and taking into account Proposition \ref{proposition}, we have proven{\footnote{The remaining details of the proof are found in Appendix 
\ref{appb}.}} the following
 \begin{theorem} \label{theorem}
 	The two-parameter first-order Lagrangian \eqref{master} for $GL(D)$-reducible independent bipartite tensors $F_{p,q}$ and $\l_{p+1,q}$, with ${\cal O}$ given by \eqref{calo} and parameters taking values in the domains $\{p\in[1,D-1],q=0\}\cup\{p\in [2,D-2],q=1\}\cup\{p=1,q\in [1,D-2]\}\cup\{p=2,q\in [2,D-3]\}$  
 	is equivalent, upon integrating out $F$ and $\l$, to two second order theories for  $GL(D)$-irreducible bipartite fields $\o_{[p-1,q]}$ and $\widehat\o_{[p',q']}$ respectively, whose field strengths are related by Hodge duality. For each of the four domains for $p$ and $q$, the values of $\{p',q'\}$ are, respectively, 
 	$\{D-p-1,0\},\, \{D-p-1,1\},\,\{D-2,q\}$ and $\{D-3,q\}$. Moreover, for the first two domains the dynamics following from each Lagrangian are dual in the sense of Section \ref{sec41}, whereas for the latter two domains this statement holds after all additional off shell fields are eliminated on shell.  
 \end{theorem}

This is in full agreement with all results in the relevant literature. The above theorem unifies all cases that have been investigated in detail so far in its realm of validity, as well as any other that would possibly be very tedious to investigate with direct methods.

\subsection{Examples}
\label{sec43}

Let us now discuss some examples and connect them to previously obtained results in the literature, as well as show how some new results can be obtained. We perform an analysis at each level of $p+q$, in other words we examine the integer partitions in two---noting that $p$ and $q$ are not interchangeable here, as should be obvious by inspection of the degree of $\l$. Here we do this for $p+q=1,2,3,4$ in order to cover a number of familiar and less known cases. Moreover, we discuss the complementary cases $p+q=D-1,D-2$, since these contain all possible additional $n$-standard duals for the fields of the previous cases, when available. 
In what follows, we initially do not include the extremal case of $p=0$, but rather comment on it collectively in the end. 

The full analysis of the cases below is covered by Proposition \ref{proposition}, Theorem \ref{theorem} and their proofs. Hence, we only present few key aspects that facilitate the comparison to previous results in the literature and highlight the universality of our approach in the sense explained in footnote \ref{foot}.

\subsubsection{p + q = 1: Scalars}
\label{sec40}

 In the present case, there are only two inequivalent partitions, only one of which is treated here, according to the comment above that we discuss the $p=0$ cases collectively at the end. 
\paragraph{Partition (p,q)=(1,0).}
In that case the ${\cal O}$ operator is trivial, namely $\mathcal{O}^{(1,0)}=\mathbb{I}$, and the parent Lagrangian \eqref{master} simply reads as
\be\label{1,0}
    \mathcal{L}^{(1,0)}_{\text{P}}=\int_{\theta,\chi}F_{1,0} \star F_{1,0}+\int_{\theta,\chi}\dd F_{1,0}\ast\widetilde\ast\,\lambda_{2,0}~.
    \ee 
    After performing the Berezin integration---note that here and in any $q=0$ case henceforth, the second integration in the second term is trivial and may be dropped in exchange for a minus sign,---this Lagrangian takes the expected component form 
    \be
    {\cal L}^{(1,0)}_{\text{P}}=-F_iF^i+\lambda_{ij}\partial^iF^j\,,
\ee
 leading 
to a $(D-2)$-form dual. Indeed, the field equation for $\l$ leads locally to 
$F_i=\partial_{i}\phi$ for a scalar field $\phi$, and on-shell \eqref{1,0} becomes its kinetic term. On the other hand, the field equation for $F$ leads to 
\be 
F_i=\frac 12\, \partial^j\l_{ij}=\frac 1{4(D-2)!}\epsilon^{i_1i_2\dots i_{D-2}}{}_{ij}\,\partial^j\,\widehat{\o}_{i_1i_2\dots i_{D-2}}~,
\ee
where the dual field $\widehat{\o}$ is defined as 
\be 
\widehat{\o}_{i_1i_2\dots i_{D-2}}:=\frac 1{2!}\epsilon_{i_1i_2\dots i_{D-2}jk}\l^{jk}~,
\ee
the Hodge dual of $\l$. Substituting this in the first order Lagrangian, the kinetic term for $\widehat{\o}$ is readily obtained, up to an irrelevant numerical factor that may be rescaled. These manipulations are easily done without reference to components, as  already proven in Section \ref{sec42}.

\subsubsection{p + q = 2: 1-forms}

In this case there are three partitions, out of which we discuss two for the time being. 
It turns out that they correspond to the standard and exotic dualization of a $1$-form gauge field. 
In both cases, the ${\cal O}$ operator is trivial. 
\paragraph{Partition (p,q)=(2,0).}
 The master Lagrangian and its corresponding component form after Berezin integration, now reads as
\be\label{2,0}
\mathcal{L}^{(2,0)}_{\text{P}}=\int_{\theta,\chi}F_{2,0} \star F_{2,0}+\int_{\theta,\chi}\dd F_{2,0}\ast\widetilde\ast\,\lambda_{3,0}=-\frac{1}{2}\,F_{ij}F^{ij}+\frac{1}{2}\,\lambda_{ijk}\partial^{i}F^{jk}\,,\ee 
which is the  master action for the standard dualization of a $1$-form, which is a $(D-3)$-form, see e.g. Eq. (2.6) of \cite{Bergshoeff}. Further details on the dualization procedure are the same as before and follow directly from the results of Section \ref{sec42}.

\paragraph{Partition (p,q)=(1,1).}

In this case we obtain a different parent Lagrangian, where the previously 2-form $F_{2,0}$ becomes a reducible bipartite (1,1) tensor $F_{1,1}$ instead and the Lagrange multiplier is a reducible bipartite $(2,1)$ tensor. Its geometric and component form respectively are
\be\label{1,1}
\mathcal{L}^{(1,1)}_{\text{P}}=\int_{\theta,\chi} F_{1,1}\star F_{1,1}+\int_{\theta,\chi}\dd F_{1,1}\ast\widetilde\ast\,\lambda_{2,1}=F_{i|j}F^{i|j}-F_{i|}{}^{i}F_{\,j|}{}^{j}+\lambda_{ij|k}\,\partial^iF^{j|k}\,,
\ee 
which is precisely the master action presented for instance in  \cite{Boulanger:2015mka} and \cite{Bergshoeff} for the exotic dualization of a $1$-form, up to an overall factor. The dual field is a $[D-2,1]$ bipartite tensor 
$\widehat\o_{[D-2,1]}$. For instance, in four dimensions, the Maxwell potential has a standard dual potential of the same type, and an exotic dual potential of type $[2,1]$.  It is very welcome that both are obtained from the same two-parameter parent Lagrangian in our formalism.

The details of the dualization procedure are a direct consequence of Proposition \ref{proposition} and Theorem \ref{theorem}. In particular, the on-shell Lagrangian after integrating out $\l$ becomes the Maxwell theory in $D$ dimensions. Up to an irrelevant overall factor, the  Lagrangian after integrating out $F$ becomes, due to Eq. \eqref{dual3},
 \be
 \mathcal{L}_{\text{dual}}(\l_{2,1})=\int_{\theta,\chi}\left(\mathbb{I}-\frac{1}{D-1}\,\eta\,\text{tr}\right)\dd^{\dagger}\l\ast\widetilde\ast\, \dd^{\dagger}\l\,, 
 \ee 
in terms of the still reducible field $\l$. Performing the Berezin integration, one easily finds that the component form of this Lagrangian is 
\be 
 \mathcal{L}_{\text{dual}}(\l_{2,1})= \partial_i\l^{ij|k}\,\partial^l\l_{lj|k}-\frac 1 {(D-1)}\,\partial_i\l^{ij|}{}_{j}\,\partial^k\l_{kl|}{}^{l}~,
\ee 
exactly what was first found in Ref. \cite{Boulanger:2015mka}. The rest of the procedure then leads, according to Eq. \eqref{eomcorrect}, to the field equation for the dual field 
\be
\text{tr}\,\dd^{\dagger}\dd\,\widehat\o_{[D-2,1]}=0~.
\ee

\subsubsection{p + q = 3: Graviton and 2-form}

In this case, the number of partitions is four, and here we present the three of them. 
Our universal Lagrangian contains the standard and exotic duals of a $2$-form gauge field, as well as the standard dual of a $[1,1]$ bipartite tensor, which is the dual graviton $[D-3,1]$. Note that the graviton also has a double standard dual $[D-3,D-3]$ \cite{Hull}. We will see that this is contained in our Lagrangian \eqref{master}, but at a different level. 

\paragraph{Partition (p,q)=(3,0).}
In this case the ${\cal O}$ operator is trivial once more, $\mathcal{O}^{(3,0)}=\mathbb{I}$. Aligning with the standard notation in the literature, we denote $F_{3,0}:= H_{3,0}$ for this example. Then the master Lagrangian  reads as
\be\label{3,0}
    \mathcal{L}^{(3,0)}_{\text{P}}=\int_{\theta,\chi} H_{3,0}\star H_{3,0}+\int_{\theta,\chi}\dd H_{3,0}\star\lambda_{4,0}=-\frac{1}{6}\,H_{ijk}H^{ijk}+\frac{1}{6}\,\lambda_{ijkl}\, \partial^iH^{jkl}\,,
    \ee 
which is precisely 
the starting point for the standard dualization of a $2$-form, the latter being a $(D-4)$-form. We do not present any further details, since there is no difference in logic to any other differential form and its standard dual.

\paragraph{Partition (p,q)=(2,1).}
For this case, in alignment with the notation in the literature, we denote $F_{2,1}:= f_{2,1}$. Unlike all previous examples, here we have to use a non-trivial ${\cal O}={\cal O}^{(2,1)}$ operator and its inverse, which according to \eqref{calo} and \eqref{inverseO} are given as \be 
\mathcal{O}=\mathbb{I}-\frac{1}{2}\,\widetilde \s\,\s~,\quad {\cal O}^{-1}=\mathbb{I}-\widetilde \s\,\s\,.
\ee 
In components, the action of this operator on the reducible field $f_{2,1}$ yields
\be\label{young 1,1}
\mathcal{O}\,f_{2,1}=\frac{1}{4}\left(f_{ijk}-f_{jki}+f_{kji}\right)\theta^i\,\theta^j\,\chi^k\,,
\ee
in accord with the index symmetry of a $(2,1)$ tensor. 
The parent Lagrangian \eqref{master} is 
\be\label{2,1}
    \mathcal{L}^{(2,1)}_{\text{P}}(f,\l)=\int_{\theta,\chi}f_{2,1}\star\mathcal{O}f_{2,1}+\int_{\theta,\chi}\dd f_{2,1}\ast\widetilde\ast\,\lambda_{3,1}~,
    \ee
    where the Lagrange multiplier is now a (reducible) mixed symmetry field of type $(3,1)$. The component form of the Lagrangian becomes
\be  \label{componentsdualgraviton}   {\cal L}^{(2,1)}_{\text{P}}(f,\l)=f_{ij|}{}^{j}f^{ik|}{}_k-\frac{1}{2}\,f_{ij|k}f^{ik|j}-\frac{1}{4}\,f_{ij|k}f^{ij|k}+\frac{1}{2}\,\lambda_{ijk|l}\partial^if^{jk|l}\,,
\ee
which is precisely the master action for the standard dualization of the graviton at the linearized level
\cite{West:2001as} (see also Eq. (2.19) of \cite{Bergshoeff} and Eq. (2.5) of \cite{Boulanger2003}). 
After solving the field equation for the Lagrange multiplier locally as
 $
f_{2,1}=\dd e_{1,1}$ or $f_{ij|k}=2\partial_{[i}e_{j]|k}, 
$
and inserting it in \eqref{componentsdualgraviton}, the result depends only on the symmetric part of the reducible field $e$. In other words, 
setting $e_{i|j}=h_{(ij)}+b_{[ij]}$, the dependence on $b$ drops out and the result is
\be\label{L12(2,1)}\begin{split}
\mathcal{L}^{(2,1)}(h)=&-h_{i|}{}^i\Box h^{j|}{}_j+2h_{i|}{}^i\partial_j\partial_kh^{j|k}+h_{i|j}\Box h^{i|j}+2\partial_ih_{j|k}\partial^kh^{i|j}\,,
\end{split}\ee
which is nothing but the linearised Einstein-Hilbert Lagrangian. This is in practice exactly what the role of ${\cal O}$ is, since the independence from the antisymmetric component $b$ would not be achieved without it. Clearly, all these manipulations are somewhat redundant in our formalism, the result being just a direct application of the general Proposition \ref{proposition}. 

Regarding the dual field,  $F$ may be integrated out using ${\cal O}^{-1}$ above, and the resulting dual second order action  in terms of the irreducible field $\widehat\o_{[D-3,1]}$ is simply the kinetic term
\be 
{\cal L}(\widehat{\o}_{[D-3,1]})=\frac 14\int_{\theta,\chi}\dd\widehat{\o}_{[D-3,1]}\star\dd\widehat{\o}_{[D-3,1]}\,.
\ee

\paragraph{Partition (p,q)=(1,2).}
In the final case at this level, ${\cal O}$ is trivial, $\mathcal{O}^{(1,2)}=\mathbb{I}$. Denoting $F_{1,2}:= Q_{1,2}$, the parent Lagrangian now reads as
\be\label{1,2}
    \mathcal{L}^{(1,2)}_{\text{P}}(Q,\l)=\int_{\theta,\chi} Q_{1,2}\star Q_{1,2}+\int_{\theta,\chi}\dd Q_{1,2}\ast\widetilde\ast\,\lambda_{2,2}~,
    \ee
    with a Lagrange multiplier being a (reducible) mixed symmetry field of type $(2,2)$.
The component form becomes 
    \be
    {\cal L}^{(1,2)}_{\text{P}}(Q,\l)=-\frac{1}{6}\,Q_{i|jk}Q^{i|jk}+\frac{1}{3}\,Q_{i|}{}^{ij}Q^{k|}{}_{kj}+\frac{1}{2}\,\lambda_{ij|kl}\partial^iQ^{j|kl}\,,
\ee
which is precisely the master action for the exotic dualization of a $2$-form  (see Eq. (3.11) of \cite{Bergshoeff}). Here we obtain this first order Lagrangian directly from \eqref{master} without the need for partial integrations used in previous works. Essentially, in our formalism the starting point is the same as for the standard dualization, albeit for different parametric values. The exotic dual is then a $[D-2,2]$ bipartite tensor, and the details follow the proof of Theorem \ref{theorem}.

Summarizing, the two-parameter universal Lagrangian we propose unifies the parent actions for the standard and exotic duals of a 2-form and the dual graviton. 

\subsubsection{p + q = 4: Curtright field and 3-form}

There are five partitions at this level. Four of them give rise to the standard and exotic duals of a $3$-form gauge field and the two different standard duals of the Curtright field $[2,1]$. The double dual of the latter is found at a different level later on.

\paragraph{Partition (p,q)=(4,0).}
This case is essentially the same as any other differential form and the parent Lagrangian \eqref{master} reduces on-shell to two dual theories for a 3-form and a 
 $(D-5)$-form dual field in the standard way.
 
\paragraph{Partition (p,q)=(3,1).}
For this partition, the operator ${\cal O}={\cal O}^{(3,1)}$ is not trivial. According to \eqref{calo} it takes the form
 \be 
 \mathcal{O}=\mathbb{I}-\frac{1}{3}\,\widetilde \s\,\s~.
 \ee 
 Its action on the reducible field $F_{3,1}$ is found to be  
\be\label{young 2,1}
\mathcal{O}\,F_{3,1}=\frac{1}{18}\left({2}F_{ijkl}+F_{ljki}+F_{klji}+F_{jkli}\right)\theta^i\,\theta^j\,\theta^k\,\chi^l\,,
\ee 
reflecting once more the component symmetry of the tensor at hand. 
The corresponding parent action is readily obtained, 
\be\label{3,1}
    \mathcal{L}_{\text{P}}^{(3,1)}(F,\l)=\int_{\theta,\chi}F_{3,1}\star\mathcal{O} 
    F_{3,1}+\int_{\theta,\chi}\dd F_{3,1}\ast\widetilde\ast\,\lambda_{4,1}~,\ee 
    with a mixed symmetry Lagrange multiplier of type $(4,1)$. The component form of this first order Lagrangian is
    \be 
 {\cal L}^{(3,1)}_{\text{P}}(F,\l)   =\frac{1}{9}\,F_{ijk|l}F^{ijk|l}+\frac{1}{6}\,F_{ijk|l}F^{ijl|k}-\frac{1}{2}\,F_{ijk|}{}^kF^{ijl|}{}_l+\frac{1}{6}\,\lambda_{ijkl|m}\partial^iF^{jkl|m}\,.
\ee
Notice that the parent actions become increasingly more complicated in component form as the degree of the field increases, containing more terms with certain relative coefficients. However, this new layer of complication is not essential in our formalism where the starting point is always the same and the dualization procedure does not change. 

For completeness and comparison to the graviton case, let us briefly explain the procedure. Solving the field equation for the Lagrange multiplier requires $F_{3,1}=\dd T_{2,1}$ locally, or in components
$F_{ijk|l}=3\partial_{[i}T_{jk]|l}$,
 with $T$ being a reducible $(2,1)$ bipartite tensor.  
 Inserting this solution in \eqref{3,1}, 
we find that if 
we decompose $T$ into its $GL(D)$-irreducible components as $T_{ij|k}=C_{ij|k}+H_{ijk}$, where $C$ is the $[2,1]$ Curtright field obeying $C_{[ij|k]}=0$ and $H$ is a 3-form, the final Lagrangian depends only on  $C_{ij|k}$. In other words, the component $H_{ijk}$ drops out (c.f. \cite{Boulanger2003}), as expected from Proposition \ref{proposition}.
Thus, on-shell the Curtright Lagrangian  is obtained, which in five dimensions reads as in Eq. \eqref{curtright1}. On the other hand, as in the graviton case, integrating out $F_{3,1}$ as detailed in Appendix \ref{appb} for the second domain, the end result is the kinetic term for the irreducible field $\widehat{\o}_{[D-4,1]}$, which is the first standard dual of the Curtright field in $D$ dimensions. Note that in five dimensions, the Curtright field is dual to the graviton, as already found in Ref. \cite{Boulanger2003}.

\paragraph{Partition (p,q)=(2,2).}
The operator ${\cal O}={\cal O}^{(2,2)}$ is again non-trivial but nevertheless different than in the previous partition, this time being \be 
\mathcal{O}=\frac 43\, \mathbb{I}-\frac{1}{3}\,\s\,\widetilde \s~.
\ee 
Its action on the reducible field $F_{2,2}$ gives 
\be 
{\cal O}\, F_{2,2}=\frac 16 \left(F_{ij|kl}-F_{il|jk}+F_{ik|jl}	\right)\theta^{i}\,\theta^j\,\chi^k\,\chi^l~.
\ee
The parent Lagrangian is obtained by substituting the values of the two parameters in \eqref{master} 
\be\label{2,2}
    \mathcal{L}^{(2,2)}_{\text{P}}(F,\l)=\int_{\theta,\chi}F_{2,2}\star \mathcal{O}F_{2,2}+\int_{\theta,\chi}\dd F_{2,2}\ast\widetilde\ast\,\lambda_{3,2}~,\ee
    the Lagrange multiplier being a $(3,2)$ reducible mixed symmetry field. 
    The corresponding component expression for the special case of $D=5$ is found in Ref. \cite{Boulanger2} 
and leads again to a proliferation of terms, which are elegantly packaged in our universal Lagrangian for any $D\ge 5$.

Dualization proceeds as explained in Section \ref{sec42} complimented with the details in Appendix \ref{appb}. Briefly, after solving the field equation for $\l$ locally as $F_{2,2}=\dd T_{1,2}$, or in components 
$ 
F_{ij|kl}=2\partial_{[i}T_{j]|kl},
$ the Lagrangian becomes on-shell the kinetic term for the Curtright field, thought of as a $[1,2]$ bipartite tensor. On the other hand, integrating out $F_{2,2}$, leads to a Lagrangian that contains the irreducible field $\widehat{\o}_{D-3,2}$, which is the second standard dual of the Curtright field. Additional fields appear in the Lagrangian too, and the duality is established upon considering the field equation \eqref{finalequationsofmotion}  for $q=2$.

\paragraph{Partition (p,q)=(1,3).}
In the final case the operator is trivial, $\mathcal{O}^{(1,3)}=\mathbb{I}$ and the master Lagrangian now reads as 
\be\label{1,3}
\mathcal{L}_{\text{P}}^{(1,3)}(F,\l)=\int_{\theta,\chi}F_{1,3}\star F_{1,3}+\int_{\theta,\chi}\dd F_{1,3}\ast\widetilde\ast\,\lambda_{2,3}~,\ee 
with a different Lagrange multiplier than before, this time of type $(2,3)$. 
The component expression is easily found to be 
\be 
{\cal L}^{(1,3)}(F,\l)=\frac{1}{6}\,F_{i|jkl}F^{i|jkl}-\frac{1}{2}\,F_{i|}{}^{ijk}F^{\,l|}{}_{ljk}+\frac{1}{6}\,\lambda_{ij|klm}\,\partial^iF^{j|klm}\,,
\ee
which is the master action for the exotic dualization of a $3$-form. We refrain from presenting any further details, since the procedure is the same as for the exotic dualization of the 2-form and it follows the general logic and the results proven in Theorem \ref{theorem}. The dual field turns out to be a 
bipartite tensor of type $[D-2,3]$. In eleven dimensions this is a $[9,3]$ field, present in the $E_{11}$ theory of \cite{West:2001as}.

\subsubsection{Double duals of the graviton and Curtright fields}

Recalling our starting aim in Section \ref{sec42}, in the presented examples we have found all 1-standard and 1-exotic duals for $p$-forms with $p=0,1,2,3$, the 1-standard dual of the linearized graviton and the two 1-standard duals of the Curtright field. Up to this level, we are still missing two duals, namely the 2-standard (double) duals of the graviton and the Curtright field. 

The missing duals are also included in the universal first order Lagrangian \eqref{master}, albeit at a different level. They are obtained as 1-standard duals of the respective 1-standard dual fields. To make this precise, let us consider the case $p+q=D-1$. From the many partitions available, we choose $p=2$ and $q=D-3$, which belongs to the fourth domain of parameter values. 
Then the Lagrangian \eqref{master} reads as 
\be\label{2,D-3}
\mathcal{L}^{(2,D-3)}_{\text{P}}(F,\l)=\int_{\theta,\chi}F_{2,D-3}\star \mathcal{O}^{(2,D-3)}F_{2,D-3}+\int_{\theta,\chi}\dd F_{2,D-3}\ast\widetilde\ast\,\lambda_{3,D-3}~,\ee
with a non-trivial ${\cal O}$. 
According to our general discussion on the dualization procedure for the fourth domain, integrating out $\l$ leads to the kinetic term for the irreducible field $[D-3,1]$, which was in turn obtained at the level $p+q=3$ as the standard dual of the graviton. 
On the other hand, using that according to \eqref{inverseo1} the inverse operator ${\cal O}^{-1}$ is 
\be 
{\cal O}^{-1}_{(2,D-3)}=\frac{D-2}{D-1}\left( \mathbb{I}+\frac{1}{2}\s\,\widetilde \s-\frac{1}{2(D-3)}\s^2\,\widetilde \s^2\right)~,
\ee 
$F_{2,D-3}$ can be integrated out. This results in a dual second order action that contains a bipartite tensor of type $[D-3,D-3]$, which is precisely the double dual of the graviton. As already explained, this action and the corresponding field equations will also contain additional fields that do not cancel algebraically. However, these fields do not propagate any degrees of freedom (in full analogy to the case of the third domain) and the duality is established upon considering the suitable trace of the field equations, in the case at hand being (see \eqref{finalequationsofmotion} with $q=D-3$)
 \be  \text{tr}^{D-3}\dd\widetilde\dd\,\widehat\o=0\,.\ee

Similarly, at  the level $p+q=D-2$ we choose the partition $p=2$ and $q=D-4$, again in the fourth domain. This leads to a first order Lagrangian with a Lagrange multiplier $\l$ being a reducible bipartite tensor $(3,D-4)$. Integrating it out leads to a second order Lagrangian for an irreducible field $[D-4,1]$, which is the first standard dual of the Curtright field found at level $p+q=4$ for $p=q=2$. On the other hand, the dual field is obtained upon integrating out $F_{2,D-4}$ and it is an irreducible bipartite tensor of type $[D-4,D-3]$. The latter is precisely the unique double dual of the Curtright field.

\subsection{Comments on the extremal case p=0 and domain walls}
\label{sec44}

Finally, we briefly discuss the case that $p=0$ in the first order Lagrangian \eqref{master}, regardless of the value of $q$. The immediate observation is that the field $F$ is a differential form of type $(0,q)$. The  zero slot is important because in the second term of \eqref{master} the operator $\dd$ acts on $F$. In other words, the field equation for $\l$ becomes 
\be \label{p0BI}
\dd\,F_{0,q}=0~. 
\ee 
This means that unlike every previous case studied in the examples, $F$ is not given locally as the differential of a field of lower degree. Instead $F$ has to be constant. 

Let us first think of the simplest case when $q=0$ too. This is the $p+q=0$ case, which we have not discussed at all up to now. Then $F$ is simply a scalar field, say $F_0$. Eq. \eqref{p0BI} means that $F_0=\text{constant}$. Moreover, since the operator ${\cal O}$ is obviously trivial, the Lagrangian \eqref{master} becomes on-shell
\be 
{\cal L}_{\text{on-shell}}^{(0,0)}(F_0)=\int_{\theta,\chi}F_0\,\star\,F_0\, \propto\,F_0^2~.
\ee
 Moreover, thinking in terms of the dualization procedure, one would obtain a dual field which would be a $(D-1)$-form and the dual Lagrangian is its standard kinetic term. 

The situation is reminiscent of what happens in ten-dimensional type IIA string theory, where the dual field is a 9-form. This is a non-dynamical field, which is not found in the spectrum of standard IIA supergravity, however it is present in IIA string theory \cite{Polchinski:1995mt} and in massive IIA supergravity \cite{Romans:1985tz}. It couples electrically to the D8-brane, which can be understood as a domain wall separating regions of different $F_0$ in massive supergravity \cite{Bergshoeff:1996ui,Chamblin:1997nnu}. 
It is in this sense that our procedure makes sense in the extremal $p=0$ case.

Following this logic, one can try to understand what happens for $q\ne 0$. These cases should also correspond to domain walls in the same spirit as above. 
In recent years it has been argued that upon toroidal string compactification and T-duality further domain wall branes exist in string theory \cite{Hassler:2013wsa,Bergshoeff:2012pm,Chatzistavrakidis:2013jqa,Chatzistavrakidis:2014sua,Bakhmatov:2017les,Otsuki:2019owg,Fernandez-Melgarejo:2018yxq}. They correspond to codimension-1 exotic states in lower dimensions.

 One example is the so-called $5_{2}^{3}$-brane, 
where the notation means a 5-brane with three additional Kaluza-Klein directions and tension scaling like $\a'^{-2}$.  Such a brane is expected to source non-geometric $R$ flux \cite{Hassler:2013wsa,Chatzistavrakidis:2014sua}. Moreover, there is evidence that this and other similar branes couple to mixed symmetry potentials, namely bipartite and multipartite tensors \cite{Bergshoeff:2011zk,Bergshoeff:2010xc,Bergshoeff:2015cba,Chatzistavrakidis:2013jqa,Chatzistavrakidis:2014sua}, a property whose origins may be found in the more general setting of the non-linear realization of the group $E_{11}$ \cite{West:2001as,Cook:2004er,West:2004kb}. In the example at hand, the $5_{2}^{3}$-brane couples to a $(9,3)$ mixed symmetry tensor. 
In our setting, this field is obtained at level $p+q=3$ with partition $p=0$ and $q=3$, i.e. the one we did not consider in the corresponding examples above. Our first order Lagrangian reads as 
\be\label{0,3}
\mathcal{L}^{(0,3)}_{\text{P}}(R,\l)=\int_{\theta,\chi}R\star R+\int_{\theta,\chi}\dd R\ast\widetilde\ast\,\lambda_{1,3}~,\ee
since ${\cal O}^{(0,3)}=\mathbb{I}$, where we denoted $F_{0,3}\equiv R$. The dualization procedure then leads to two dual second order Lagrangians, the first being simply 
\be 
\int_{\theta,\chi}R\star R\,\propto R^{ijk}R_{ijk}~,
\ee
for $R^{ijk}=\text{constant}$. This is then the analogon of the mass deformation, but this time with a constant $(0,3)$ bipartite tensor, which may also be viewed as a trivector associated to a constant $R$ flux magnetically sourced by the $5_{2}^{3}$-brane. The dual theory then involves a $[9,3]$ 
bipartite tensor, the analogon of the $9$-form in IIA string theory.

The above comments provide a basis for understanding the $p=0$ case, which should then indeed be allowed in \eqref{master}. We leave a more complete treatment for these fields and the corresponding sources for future work.

\paragraph{Acknowledgements.}  We would like to thank Clay Grewcoe for helpful comments and Larisa Jonke for useful discussions and suggestions on the manuscript. The work of A.Ch. and G.K. is supported by the Croatian Science Foundation Project ``New Geometries for Gravity and Spacetime'' (IP-2018-01-7615), and also partially supported by the European Union through the European Regional Development Fund - The Competitiveness and Cohesion Operational Programme (KK.01.1.1.06). 

\appendix

\section{Useful identities and additional proofs}
\label{appa} 

In this appendix we collect some formulas used in the main text. First, we present some relations among the maps defined in Section \ref{sec2}, which appear in Refs. \cite{Bekaert:2002dt,deMedeirosHull1,Bekaert:2003az,deMedeirosHull2,deMedeiros:2003qel} in a slightly different 
setting. Although graded geometry is not used in those works, the following formulas still hold for maps between bipartite tensors, since in local coordinates our definitions are completely equivalent to the ones proven in the aforementioned papers. Alternatively, they may be easily proven using \eqref{alt} and the following definitions of ``number operators'' 
\be 
\hat{p}=\theta^i\bar{\theta}_i \quad \text{and} \quad \hat{q}=\chi^i\bar{\chi}_i\,,
\ee which count the degrees $p$ and $q$ respectively of a bipartite tensor of type (p,q). Then the set of six operators $(\eta, \text{tr},\s,\widetilde{\s},\hat{p},\hat{q})$ satisfy the commutation relations
\be \begin{split}\label{so4}
	&[\s,\widetilde{\s}]=\hat{p}-\hat{q}\,,\quad[\text{tr},\h]=D-\hat{p}-\hat{q}\,,
\quad	[\hat{p},\s]=-[\hat{q},\s]=\s\,,\\[4pt] &[\hat{p},\widetilde{\s}]=-[\hat{q},\widetilde{\s}]=-\widetilde{\s}\,,\quad
	[\hat{p},\text{tr}]=[\hat{q},\text{tr}]=-\text{tr}\,,\quad[\hat{p},\h]=[\hat q,\h]=\h\,,\\[4pt]
	&[\s,\h]=[\widetilde{\s},\h]=[\s,\text{tr}]=[\widetilde{\s},\text{tr}]=0\,.
\end{split}
\ee We note in passing that these six maps and their commutation relations may be related to the six generators $J_{ab}\,, a,b=1,2,3,4$ of the Lie algebra $\mf{so}(4)$, see \cite{Bastianelli:2007pv,Bonezzi:2018box}, as \be\begin{split}
	&J_{13}=-\hat p\,,\quad J_{24}=-\hat q\,,\\[4pt]& J_{12}= -\frac{i}{2}(\s-\widetilde{\s}+\h-\text{tr})\,, \quad J_{14}=-\frac{1}{2}(\s+\widetilde{\s}-\h-\text{tr})\,,\\
	&J_{23}=-\frac{1}{2}(\h+\text{tr}+\s+\widetilde{\s})\,, \quad J_{34}=-\frac{i}{2}(\h-\text{tr}-\s+\widetilde{\s})\,,
\end{split}
\ee 
where 
\be 
[J_{ab},J_{cd}]=4i\d_{[a[c}J_{b]d]}\,.
\ee 
This $\mf{so}(4)$ is generated by the Poisson algebra of the associated Poisson bracket that can be defined for the graded coordinates and momenta of the graded phase space mentioned in Section \ref{sec21}. 
Identities containing (co-)differentials may be worked out in the same way. The ones we use in this paper are summarized as 
\bse\label{ids}\begin{align}
\label{id2}& \dd\,\s+\s\,\dd=0~,\\[3pt]
&\dd\,\eta+\eta\,\dd=0~,\\[3pt]
& \label{id1}\dd^{\dagger}\,\text{tr}+\text{tr}\,\dd^{\dagger}=0~,\\[3pt]
\label{id3}& \dd\,\widetilde \s^{n}+(-1)^{n+1}\widetilde \s^n\,\dd=-n\, \widetilde\dd\,\widetilde \s^{n-1}~,\\[3pt]
&\label{id7} \dd\,\text{tr}^n+(-1)^{n+1}\text{tr}^n\,\dd=n\,\widetilde\dd^{\dagger}\,\text{tr}^{n-1}~,\\[3pt]
&\label{id10}\dd^{\dagger}\,\eta^{n}+(-1)^{n+1}\eta^{n}\,\dd^{\dagger}=n\,\widetilde\dd\,\eta^{n-1}~.
\end{align}\ese
Note that there are also the corresponding identities for transposed maps.  It is worth noting that the quadruple $(\dd,\widetilde{\dd},\dd^{\dagger},\widetilde{\dd}^{\dagger})$ may be related to four  supercharges $Q_{a}$ as 
\be \begin{split}
	&Q_1\equiv \frac{-i}{\sqrt{2}}(\dd+\dd^\dagger)\,,\quad Q_2\equiv \frac{-i}{\sqrt{2}}(\widetilde{\dd}+\widetilde{\dd}^\dagger)\,,\quad
	Q_3\equiv \frac{1}{\sqrt{2}}(\dd-\dd^\dagger)\,,\quad Q_4\equiv \frac{1}{\sqrt{2}}(\widetilde{\dd}-\widetilde{\dd}^\dagger)\,,
\end{split}
\ee 
satisfying 
\be 
Q_{a}^2=H \quad \text{and}\quad [Q_{a},H]=0\,,
\ee 
where $H$ is the bosonic operator $-\frac{\Box}2$ (the Hamiltonian) \cite{Bonezzi:2018box}. In other words they generate a ${\cal N}=4$ supersymmetry algebra and the $\mf{so}(4)$ algebra is its R-symmetry, as can be seen through \eqref{ids}.{\footnote{Our sign conventions differ slightly from standard literature in that while $\{Q_1,Q_3\}=\{Q_2,Q_4\}=0$, there are also commutation relations, e.g. $[Q_1,Q_2]=[Q_1,Q_4]=[Q_2,Q_3]=[Q_3,Q_4]=0$. Certainly all these become anticommutators in the opposite sign convention. Our chosen convention is more practical in the context of the main body of the paper though. }} Here we do not utilize further the relation to supersymmetric mechanics, but only use the full set of identities \eqref{so4} and \eqref{ids}.

Moreover, a number of integral identities hold for all bipartite tensor fields with degree as indicated. One of them is
\be
\label{integralidentity2}
\int_{\theta,\chi}\o_{p,q}\, \xi_{D-p,D-q}=-\int_{\theta,\chi}\ast\o_{p,q}\ast\xi_{D-p,D-q}=-\int_{\theta,\chi}\widetilde\ast\,\o_{p,q}\,\widetilde\ast\,\xi_{D-p,D-q}\,,
\ee
which can be easily proven using the definitions \eqref{ast1} and \eqref{ast2}. A second identity  we use appears in \cite{deMedeiros:2003qel} and reads as
\be\label{integralidentity3}
\int_{\theta,\chi}\o_{p,q}\ast\widetilde\ast\, \eta\, \xi_{p-1,q-1}=\int_{\theta,\chi}\text{tr}\,\o_{p,q}\ast\widetilde\ast\,\xi_{p-1,q-1}\,.
\ee
Finally, the identities 
\be \label{integralidentity1}
\begin{split}
\int_{\theta,\chi}\widetilde \s^n\s^n\ast\o_{p,q}\,\widetilde\ast\, \xi_{p,q}&=(-1)^{p(D-p)}\int_{\theta,\chi}\eta^n\,\text{tr}^n\o_{p,q}\ast\widetilde\ast\,\xi_{p,q}\,,\\
\int_{\theta,\chi}\s^n\widetilde \s^n\,\widetilde\ast\,\o_{p,q}\ast \xi_{p,q}&=(-1)^{q(D-q)}\int_{\theta,\chi}\eta^n\,\text{tr}^n\o_{p,q}\ast\widetilde\ast\,\xi_{p,q}\,,
\end{split}\ee
also hold for any $n$. We will now prove the first of these identities for $n=1$; the other cases, as well as the second identity, can be proven in exactly the same manner. Using the definitions of the maps appearing in the left hand side, we find
\bea
\int_{\theta,\chi}\widetilde \s\,\s\ast\o_{p,q}\,\widetilde\ast\, \xi_{p,q}&=&(-1)^{qD+p+1}\int_{\theta,\chi}\widetilde\ast(\text{tr}\ast\widetilde\ast\,\text{tr}\,\o_{p,q})\,\widetilde\ast\,\xi_{p,q}\nn\\[3pt]
&\overset{\eqref{integralidentity2}}=&(-1)^{qD+p}\int_{\theta,\chi}\text{tr}(\ast\widetilde\ast\,\text{tr}\,\o_{p,q})\,\xi_{p,q}\nn\\[3pt]
&=&(-1)^{pD+q}\int_{\theta,\chi}\text{tr}(\ast\widetilde\ast\,\text{tr}\,\o_{p,q})\ast\widetilde\ast(\ast\widetilde\ast\,\xi_{p,q})\nn\\[3pt]
&\overset{\eqref{integralidentity3}}=&(-1)^{pD+q}\int_{\theta,\chi}\ast\widetilde\ast\,\text{tr}\,\o_{p,q}\ast\widetilde\ast(\eta\ast\widetilde\ast\,\xi_{p,q})\nn\\[3pt]
&\overset{\eqref{integralidentity2}}=&(-1)^{pD+q}\int_{\theta,\chi}\text{tr}\,\o_{p,q}\,\eta\ast\widetilde\ast\,\xi_{p,q}\nn\\[3pt]
&=&(-1)^{p(D-p)}\int_{\theta,\chi}\eta\,\text{tr}\,\o_{p,q}\ast\widetilde\ast\,\xi_{p,q}\,,
\eea
which is indeed the first identity in \eqref{integralidentity3} for $n=1$. 

\paragraph{Cyclicity of the integral for the $\star$ operator.} 
  Now we proceed in proving that 
  \bea \label{cyclicity}
  \int_{\theta,\chi}\o_{p,q}\star\xi_{p,q}=\int_{\theta,\chi}\xi_{p,q}\star\o_{p,q}~.
  \eea 
   First, we note that integrals of this form are obviously invariant under the transposition operator $\top$ (tilde operation) since they correspond to spacetime scalars. Thus, we have
 \bea 
 \int_{\theta,\chi}\o_{p,q}\star\xi_{p,q}&=&\left(\int_{\theta,\chi}\o_{p,q}\star\xi_{p,q}\right)^\top= \int_{\theta,\chi}\o^\top_{q,p}(\star\,\xi_{p,q})^\top =\nn\\[3pt]
 							&\overset{\eta^\top=\,\eta}=&\int_{\theta,\chi}\o^\top_{q,p}\frac{1}{(D-p-q)!}\,\eta^{D-p-q}(\xi_{p,q}^\top)^\top= \nn\\[3pt]
 								&\overset{(\xi^\top)^\top=\,\xi}=&\int_{\theta,\chi}\xi_{p,q}\frac{1}{(D-p-q)!}\,\eta^{D-p-q}\o_{q,p}^\top =\int_{\theta,\chi}\xi_{p,q}\star\o_{p,q}~,
 \eea 
 which proves \eqref{cyclicity}.
 
 \paragraph{Proof of Lemma \ref{Poincarelemma}.} 
 We give a proof that is valid for an arbitrary smooth manifold and
repeatedly uses the ordinary Poincar\'e lemma, which holds in either of the odd variables $\theta^i$ and $\chi^i$ separately. On a contractible patch, $\dd \widetilde\dd \,\xi_{p,q} =
 \dd(\widetilde\dd \,\xi_{p,q}) = 0$ implies 
 \be 
 \widetilde\dd\, \xi_{p,q} = \dd \,\xi_{p-1, q+1}~, \label{chain}
 \ee
 for some new mixed symmetry tensor $\xi_{p-1, q+1}$. Since $(\widetilde\dd)^2 = 0$,  this in turn implies $\dd \widetilde\dd \,\xi_{p-1, q+1} = 0$. Vice versa, $\dd \widetilde\dd \,\xi_{p-1, q+1} = 0$ implies \eqref{chain} and $\dd \widetilde\dd \,\xi_{p,q}=0$. We can use this chain of relations with increasing and decreasing degrees in the odd variables to prove the lemma by induction. With the conventions explained above, let us start with the induction hypothesis
 \be
 \xi_{p-1,q+1} = \dd \kappa_{p-2,q+1} + \widetilde\dd \kappa_{p-1,q} + c_{i_1 \ldots i_p k_0 \ldots k_q} \theta^{i_1} \ldots \theta^{i_{p-1}} x^{i_p} \chi^{k_0} \ldots \chi^{k_q}~.
 \ee
 Using \eqref{chain}, this implies
 \be
 \widetilde\dd\, \xi_{p,q} = 0 + \dd \widetilde\dd \kappa_{p-1,q} + c_{i_1 \ldots i_p k_0 \ldots k_q} \theta^{i_1} \ldots \theta^{i_{p}} \chi^{k_0} \ldots \chi^{k_q}\,,
 \ee
 which can be rewritten to obtain
 \be
 \widetilde\dd(\xi_{p,q} - \dd \kappa_{p-1,q} - c_{i_1 \ldots i_p k_0 \ldots k_q} \theta^{i_1} \ldots \theta^{i_{p}} x^{k_0} \chi^{k_1} \ldots \chi^{k_q}) = 0\,, 
 \ee
 and hence
 \be
 \xi_{p,q} - \dd \kappa_{p-1,q} - c_{i_1 \ldots i_p k_0 k_1\ldots k_q} \theta^{i_1} \ldots \theta^{i_p} x^{k_0} \chi^{k_1} \ldots \chi^{k_q} =
 \widetilde\dd\, \kappa_{p,q-1} \,,
 \ee
 i.e. \eqref{lemma} for some mixed symmetry tensor $\kappa_{p,q-1}$. 
 
 The suitable choice of induction anchor depends on the value of $p+q$:
 
 For $p+q \geq D$ we can use the chain of relations until we reach $\dd \widetilde\dd\, \xi_{p+q-D, D}=0$. Clearly, $\widetilde\dd (\xi_{p+q-D, D} - \dd \kappa_{p+q-D-1,D}) = 0$ (both terms are of maximal degree in~$\chi^k$) and hence  $\xi_{p+q-D, D} = \dd \kappa_{p+q-D-1,D} + \widetilde\dd \kappa_{p+q-D, D-1}$. (The first term can in fact be absorbed in the second one, but we write the expression in this way to match the lemma that we would like to prove.) Starting from this anchor, we can use the induction step by step to prove the lemma for the case $p+q \geq D$ without $c$-term, as desired.
 
 For $p+q < D$ we can use the chain of relations until we reach $\dd \widetilde\dd\, \xi_{0,p+q}=0$. This implies $\widetilde\dd\, \xi_{0,p+q} = \text{const.} =: c_{k_0 \ldots k_{p+q}} \chi^{k_0} \ldots \chi^{k_{p+q}}$, where $c_{k_0 \ldots k_{p+q}}$ is a totally antisymmetric constant tensor,  and rearranging slightly 
 \be
 \widetilde\dd(\xi_{0,p+q} - c_{k_0 k_1 \ldots k_{p+q}} x^{k_0} \chi^{k_1}  \ldots \chi^{k_{p+q}}) =0 \,,
 \ee
 hence, using Poincar\'e lemma in $\chi^k$
 \be
 \xi_{0,p+q} = \widetilde\dd \kappa_{0,p+q-1} + c_{k_0 k_1 \ldots k_{p+q}} x^{k_0} \chi^{k_1}  \ldots \chi^{k_{p+q}} \,.
 \ee
 This is the desired anchor for the proof by induction in the case  $p+q < D$.

\paragraph{Proof of Eq. \eqref{app1}.} 
Using the map identities \eqref{id2} and \eqref{id3}, we compute
\bea 
\dd\,\s^n\,\widetilde \s^n&=&(-1)^n\,\s^n\,\dd\,\widetilde \s^n \nn\\[3pt] 
							&=& (-1)^n\,\s^n\left(-(-1)^{n+1}\,\widetilde \s^n\,\dd-n\,\widetilde\dd\,\widetilde \s^{n-1}\right)\nn\\[3pt]
							&=&\s^n\,\widetilde \s^{n}\,\dd+(-1)^{n+1}n\,\s^n\,\widetilde\dd\,\widetilde \s^{n-1} \nn\\[3pt] 
							&=&\s^n\,\widetilde \s^{n}\,\dd+(-1)^{n+1}n\left((-1)^n\widetilde\dd\,\s^n+(-1)^nn\,\dd\,\s^{n-1}\right)\widetilde \s^{n-1} \nn\\[3pt]
							&=&\s^n\,\widetilde \s^{n}\,\dd+\widetilde\dd Y-n^2\dd\,\s^{n-1}\,\widetilde \s^{n-1}~,
\eea 
which directly leads to the recursive formula
\be 
\dd\left(\s^n\,\widetilde \s^n+n^2\,\s^{n-1}\,\widetilde \s^{n-1}\right)\o=\s^n \,\widetilde \s^n\dd\o+\widetilde\dd Y~,\nn
\ee 
as desired. Here $Y=-n\,\s^n\,\widetilde \s^{n-1}\o$, although it is unnecessary to present explicitly such quantities for our purposes, since they always do not influence the results.

\paragraph{Proof of Eq. \eqref{app2}.}
First we define for brevity 
\be 
P(n):=\dd\,\s^n\,\widetilde \s^n \quad \text{and} \quad Q(n)=\s^n\,\widetilde \s^n\,\dd~.
\ee
Then we write \eqref{app1} for every $n$ as follows 
\bea 
P(n)+n^2P(n-1)&=&Q(n)+\widetilde\dd Y_1 \nn\\[3pt]
-\,n^2\left(P(n-1)+(n-1)^2P(n-2)\right)&=&-\,n^2Q(n-1)+\widetilde\dd Y_2 \nn\\[3pt]
+\,n^2(n-1)^2\left(P(n-2)+(n-2)^2P(n-3)\right)&=&+\,n^2(n-1)^2Q(n-2)+\widetilde\dd Y_3 \nn\\[3pt]
\vdots \qquad \qquad\qquad &=& \qquad \qquad \qquad \vdots \nn\\[3pt]
+\,(-1)^{n-1}n^2(n-1)^2\dots2^2\left(P(1)+P(0)\right) &=&+\,(-1)^{n-1}n^2(n-1)^2\dots2^2 Q(1)+\widetilde\dd Y_{n}\nn\\[3pt]
+\,(-1)^n(n!)^2 \,P(0)&=&+\,(-1)^n(n!)^2 \, Q(0)+\widetilde\dd Y_{n+1}~.\nn
\eea 
 Summing up these $n+1$ equations, the right hand side yields the sought-after $P(n)$, which turns out to be 
 \be 
 P(n)=Q(n)+\sum_{k=1}^n(-1)^k\prod_{m=0}^{k-1}(n-m)^2\,Q(n-k)+\widetilde\dd Y'~, 
 \ee 
 for some calculable $Y'$, which is precisely the relation \eqref{app2}.

 \section{Proof of Theorem \ref{theorem}}
 \label{appb}
 
 In this second appendix, we provide the missing technical details for the proof of Theorem \ref{theorem}. The part that refers to the Proposition \ref{proposition} has already been proven in the main text in full detail. Hence we concentrate in showing how the correct field equations and degrees of freedom for the dual field are obtained in each of the remaining three out of four domains of $p$ and $q$ values of the two-parameter parent Lagrangian. (Recall that the simplest case of the first domain was already fully proven in the main text.)

In what follows, we vary the parent Lagrangian \eqref{master} w.r.t. the field $F_{p,q}$. This variation is not trivial; we  first show that
\be\label{variationintegral}
\int_{\theta,\chi}\d\left(F\star\mathcal{O}F\right)=\int_{\theta,\chi}\d F\star\mathcal{O}F+\int_{\theta,\chi}F\star\mathcal{O}\d F=2\int_{\theta,\chi}\d F\star\mathcal{O}F
\ee 
for any $F_{p,q}$ and $\mathcal{O}^{(p,q)}$ given by \eqref{calo}. Acting on any bipartite tensor of type $(p,q)$,  the general identities 
\be\label{identitiessigmatrace}
\text{tr}^n\ast=(-1)^{n(p+1)}\ast\s^n\,\,,\qquad\text{tr}^n\widetilde\ast=(-1)^{n(q+1)}\,\widetilde\ast\,\widetilde \s^n
\ee
 follow directly from definition \eqref{sigma maps} and are valid for every $n\geq 1$. Since the form of $\mathcal{O}$ depends on whether $p\geq q+1$ or $p<q+1$, \eqref{variationintegral} must be proven in both  cases. We  refrain from writing the proof for the second case, since all necessary steps and technical manipulations are completely analogous to the first one.

The first step is to observe that $\star\,\mathcal{O}=\mathcal{O}\,\star$ by definition, since both $\s$ and $\widetilde \s$ maps contained in $\mathcal{O}$ commute with both $\eta$ and $\text{tr}$ maps contained in $\star$. Thus, for $p\geq q+1$ one computes
\be\begin{split}
\int_{\theta,\chi}F\star\mathcal{O}\d F&=\int_{\theta,\chi}F\mathcal{O}(\star\d F)=\int_{\theta,\chi}\mathcal{O}\left[\ast(\ast\star\d F)\right]\widetilde \ast(\widetilde \ast F)\\
&\overset{\eqref{calo}}{=}\int_{\theta,\chi}\d F\star F+\sum_{n=1}^qc_n\int_{\theta,\chi}\widetilde \s^n\s^n\ast(\ast\star \d F)\widetilde \ast(\widetilde \ast F)\\
&\overset{\eqref{integralidentity1}}{=}\int_{\theta,\chi}\d F\star F+(-1)^{p(D-p)}\sum_{n=1}^qc_n\int_{\theta,\chi}\eta^n\text{tr}^n(\ast\star \d F)\ast\widetilde\ast(\widetilde\ast F)\\
&\overset{\eqref{integralidentity3}}{=}\int_{\theta,\chi}\d F\star F+(-1)^{p(D-p)}\sum_{n=1}^qc_n\int_{\theta,\chi}\ast\star \d F\ast \widetilde\ast (\eta^n\text{tr}^n\widetilde\ast F)\,.
\end{split}
\ee
Using the second identity in \eqref{identitiessigmatrace}, \eqref{integralidentity2} and \eqref{integralidentity3} one gets 
\begin{equation*}
\int_{\theta,\chi}F\star\mathcal{O}\d F=\int_{\theta,\chi}\d F\star F+(-1)^{q(D-q)}\sum_{n=1}^q
c_n(-1)^{np}\int_{\theta,\chi}\widetilde\ast\star\d F\ast\widetilde\ast (\text{tr}^n\ast\widetilde \s^nF)\,.
\end{equation*}
Using now the first identity in \eqref{identitiessigmatrace}, \eqref{integralidentity2} and the integral cyclicity \eqref{cyclicity} we have
\begin{equation}\begin{split}
\int_{\theta,\chi}F\star\mathcal{O}\d F&=\int_{\theta,\chi}\d F\star F+\sum_{n=1}^qc_n\int_{\theta,\chi}\s^n\widetilde \s^nF\star\d F\\
&=\int_{\theta,\chi}\mathcal{O}F\star\d F=\int_{\theta,\chi}\d F\star\mathcal{O}F\,,
\end{split}
\end{equation}
which proves \eqref{variationintegral} for the $p\geq q+1$ case. We are now ready to detail the proof of Theorem \eqref{theorem} for the rest of  parameter domains.

\paragraph{The second domain.} The second domain is the one for which $\{p\in[2,D-2], q=1\}$ and it corresponds to the standard dualization of generalized graviton fields. In this case, the parent Lagrangian reads as
\be\label{b1}
\mathcal{L}_{\text{P}}=\int_{\theta,\chi}F_{p,1}\star\mathcal{O}^{(p,1)}F_{p,1}+\int_{\theta,\chi}\dd F_{p,1}\ast\widetilde\ast\,\l_{p+1,1}~;
\ee
 varying w.r.t. to $F$ and taking into account \eqref{variationintegral}, the following duality relation is readily obtained
\be\label{b2}
\star\mathcal{O}F=\frac{(-1)^{p+1}}{2}\,\widetilde\ast\,\dd\ast\l\,.
\ee  
Firstly, we will show that the on-shell version of this relation leads to an equation of the form \eqref{dr2}. More precisely, we will assume the equation $\dd F=0$, which comes from varying \eqref{b1} w.r.t. $\l$. This is equivalent to demanding that locally $F=\dd\o$, for a reducible field $\o_{p-1,1}$. According to the defining property \eqref{calorequirement} of $\mathcal{O}$, we get
\be
\star \dd [\o]=\frac{(-1)^{p+1}}{2}\,\widetilde{\ast}\,\dd\ast\l-\star\,\widetilde{\dd}X,
\ee
where $[\o]:=\o_{[p-1,1]}$ and $X:=X_p$ is an arbitrary $p$-form. Using \eqref{starrelations}, the above relation takes the form
\be \label{relation RR}
\dd[\o]-\h\,\text{tr}\dd[\o]=\frac{(-1)^{\e(p,1)+1}}{2}\,\dd^\dagger\l-\widetilde{\dd}X+\h\,\text{tr}\widetilde{\dd}\,X
\ee
Tracing both sides of this relation will give 
\be 
\text{tr}\dd[\o]=\frac{(-1)^{\e(p,1)}}{2(D-p)}\,\text{tr}\dd^\dagger\l-\text{tr}\widetilde{\dd}X
\ee
and, by substituting this back into \eqref{relation RR}, one finds 
\be \label{onshellDR}
\dd[\o]=\frac{(-1)^{\e(p,1)+1}}{2}\left(\mathbb{I}-\frac{1}{D-p}\,\h\,\text{tr}\right)\dd^\dagger\l-\widetilde{\dd}X\,.
\ee
This is the on-shell duality relation. Decomposing $\l$ as in \eqref{decomp1} and acting on \eqref{onshellDR} with $\widetilde{\dd}$, a little algebra  leads to the relation \eqref{dr2}. 
\color{black}
\\
Let us now proceed in finding the dual action. Applying \eqref{starrelations} to ${\cal O}F$, 
relation \eqref{b2} may be rewritten as
\be\label{b3}
\mathcal{O}(F-\eta\,\text{tr}F)=\frac{(-1)^{\e(p,1)+1}}{2}\,\dd^\dagger\l\,,
\ee
where we used that $\mathcal{O}$  contains only the  maps $\s$ and $\widetilde \s$, which commute with both $\eta$ and $\text{tr}$ by virtue of identities \eqref{so4}. 
Next, the duality relation should be solved for $F$, which requires the inverse operator $\mathcal{O}^{-1}$, 
namely the one that satisfies $\mathcal{O}^{-1}\mathcal{O}=\mathcal{O}\mathcal{O}^{-1}=\mathbb{I}$ on any bipartite tensor. Rather than proving the invertibility of ${\cal O}$ abstractly, we construct  its inverse using direct methods. In the present case, $p\geq q+1=2$ and one finds
\be\label{inverseO}
(\mathcal{O}^{(p,1)})^{-1}=\mathbb{I}-\widetilde \s\,\s\,.
\ee
Acting with it on both sides of \eqref{b3} one gets
\be\label{0.7}
F-\eta\,\text{tr}F=\frac{(-1)^{\e(p,1)+1}}{2}(\mathbb{I}-\widetilde \s\s)\,\dd^\dagger\l\,.
\ee 
Next, taking  the trace of \eqref{0.7} and using the second commutation relation in \eqref{so4}, one can determine $\text{tr}F_{p,1}$ in terms of $\l_{p+1,1}$. Inserting it back in \eqref{0.7} one finds \be\label{0.8}
F=\frac{(-1)^{\e(p,1)+1}}{2}\left(\mathbb{I}-\widetilde \s\s-\frac{1}{D-p}\,\eta\,\text{tr}\right)\dd^\dagger\l\,.
\ee 
Substituting  \eqref{b2} and \eqref{0.8} into \eqref{b1},  the dual Lagrangian \eqref{-dual1} in terms of $\l$ is obtained as
\be\label{dual1}
\mathcal{L}_{\text{dual}}(\l)=\frac{(-1)^{\e(p,1)+1}}{4}\int_{\theta,\chi}\left(\mathbb{I}-\widetilde \s\s-\frac{1}{D-p}\,\eta\,\text{tr}\right)\dd^\dagger\l\ast\widetilde\ast\,\dd^\dagger \l\,.
\ee 
 Now we decompose the Lagrange multiplier as in \eqref{decomp1} and introduce the two independent fields $\widehat\o_{[D-p-1,1]}$ and $\mathring{\l}_{p,0}$. Using the second of \eqref{so4}, \eqref{id3} and \eqref{id10} one can easily show that the first factor in the above integrand only depends on $\widehat\o$, i.e.
\bea
\mathcal{L}_{\text{dual}}(\widehat{\o},\mathring{\l})&=&\frac{(-1)^{\e(p,1)+1}}{4}\int_{\theta,\chi}\left\{\left(\mathbb{I}-\widetilde \s\s-\frac{1}{D-p}\,\eta\,\text{tr}\right)\dd^\dagger\ast\widehat{\o}\right\}\ast\widetilde\ast\,\dd^\dagger (\ast\,\widehat{\o}+\eta\mathring{\l})\nn\\
&=&\frac{(-1)^{\e(p,1)+1}}{4}\int_{\theta,\chi}\left\{\left(\mathbb{I}-\widetilde \s\s\right)\dd^\dagger\ast\widehat{\o}\right\}\ast\widetilde\ast\,\dd^\dagger (\ast\,\widehat{\o}+\eta\mathring{\l})\,,
\eea
where in the second line we used  identity \eqref{id1} and  that  $\text{tr}\ast\widehat{\o}_{[D-p-1,1]}=0$ due to the tracelessness of $\widehat{\l}$. Finally, using the integral identity \eqref{integralidentity3}, the identities \eqref{id7} - \eqref{id10} and the definitions \eqref{sigma maps} - \eqref{codifferentials}, one finds that the $\mathring{\l}$-dependence in the dual Lagrangian cancels algebraically. The result is then
\be\label{dual1234}
\begin{split}
\mathcal{L}_{\text{dual}}(\widehat{\o})
&=\frac{(-1)^{\e(p,1)+p(D-p)}}{4}\int_{\theta,\chi}\left(\mathbb{I}-\widetilde \s\s\right)\ast\dd\,\widehat{\o}\,\widetilde\ast\,\dd\,\widehat{\o}\\
&=\frac{(-1)^{\e(p,1)}}{4}\int_{\theta,\chi}\dd\,\widehat{\o}\ast\widetilde\ast\left(\dd\,\widehat{\o}-\eta\,\text{tr}\,\dd\,\widehat{\o}\right)=\frac 14\int_{\theta,\chi}\dd\,\widehat{\o}\star\dd\,\widehat{\o}\,,
\end{split}
\ee
where the three integral identities \eqref{integralidentity2}, \eqref{integralidentity3} and \eqref{integralidentity1} were used. 
Finally, we note that the original field $\o_{[p-1,1]}$ and the dual field $\widehat\o_{[D-p-1,1]}$ have the same number of components as $SO(D-2)$ representations, since 
\small\bea 
\binom{D-2}{p-1}\binom{D-1}{1}(1-\frac{1}{p})-\binom{D-2}{p-2}=\binom{D-2}{D-p-1}\binom{D-1}{1}(1-\frac{1}{D-p})-\binom{D-2}{D-p-2}\,.\nn\\
\eea
\normalsize
 This implies directly that in the physical gauge the number of field equations for the two fields is the same. As discussed in the main text, this establishes Theorem \ref{theorem} for this domain of parameters. 

\paragraph{The third domain.} Now $\{p=1, q\in[1,D-2]\}$ and this domain corresponds to the exotic dualization of a $q$-form field. The parent Lagrangian reads in this case as
\be\label{b11}
\mathcal{L}_{\text{P}}=\int_{\theta,\chi}F_{1,q}\star F_{1,q}+\int_{\theta,\chi}\dd F_{1,q}\ast\widetilde\ast\l_{2,q}
\ee
and the duality relation obtained from the variation w.r.t. $F$ can be brought to the form
\be\label{DR third domain}
F=\frac{(-1)^{\e}}{2}\left(\mathbb{I}-\frac{1}{D-q}\,\eta\,\text{tr}\right)\dd^\dagger\l\,.
\ee 
One can easily prove that the on-shell version of this relation leads to  \eqref{dr3}. First, we decompose $\l$ as in \eqref{dec} and consider that locally $F=\dd\o$,  $\omega_{0,q}$ being a $q$-form. Acting with $\widetilde{\dd}$ on the resulting relation will give
\be 
2(-1)^{\e(1,q)}\widetilde{\dd}\,\dd\,\o=\widetilde{\dd}\dd^\dagger\ast\widehat\o+\widetilde{\dd}\dd^\dagger\h\mathring{\l}+\frac{1}{D-q}\,\widetilde{\dd}\,\h\,\dd^\dagger\text{tr}\,\h\mathring{\l}\,.
\ee 
The $\mathring{\l}$-dependent terms cancel each other, as can be seen by using suitable identities from Appendix \ref{appa}. Thus, the on-shell duality relation reads as
\be 
\widetilde{\dd}\,\dd\,\o=\frac{(-1)^{qD}}{2}\,\ast\,\widetilde{\dd}\,\dd\,\widehat\o\,,
\ee
which is precisely \eqref{dr3}.

 The dual Lagrangian \eqref{dual3} in terms of $\l$ is obtained by substitution of $F$, 
\be\label{dualthree}
\mathcal{L}_{\text{dual}}[\l]=\frac{3(-1)^{\e}}{8}\int_{\theta,\chi}\left(\mathbb{I}-\frac{1}{D-q}\eta\,\text{tr}\right)\dd^\dagger\l\ast\widetilde\ast\,\dd^\dagger\l\,.
\ee 
 In contrast to the previous case, this Lagrangian depends explicitly  on the trace part $\mathring{\l}_{1,q-1}$ of the Lagrange multiplier. However, one should obtain  equations of motion only for the dual field $\widehat{\o}_{[D-2,q]}$ such that the duality of the two theories is justified, as explained in the main text. Varying \eqref{dualthree} w.r.t. $\l$ one obtains
\be\label{eomseoms}
\dd\left(\mathbb{I}-\frac{1}{D-q}\,\eta\,\text{tr}\right)\dd^{\dagger}\l=0\,.
\ee  
After the decomposition \eqref{dec}, and using the fact that $\text{tr}\ast\dd\widehat{\o}_{[D-2,q]}\propto\ast\dd\s\, \widehat{\o}_{[D-2,q]}=0$, which follows from \eqref{sigma maps} and \eqref{id2}, one obtains
\be\label{b155}
\dd\ast\dd\widehat\o-\dd\dd^\dagger\eta\mathring{\l}+\frac{1}{D-q}\dd\eta\,\text{tr}\,\dd^\dagger\eta\mathring{\l}=0\,.
\ee
Acting now with $\text{tr}^q\ast$ on both sides, it is observed that the second and  third terms vanish due to the first identity in \eqref{identitiessigmatrace}. Then the second term in \eqref{b155} is proportional to $\dd\s^q\dd^\dagger\eta\mathring{\l}_{1,q-1}$, which vanishes as may be shown using the identities \eqref{id10}, \eqref{id3} and \eqref{so4}. For the third term in \eqref{b155}, one can see that it is proportional to $\s^q\,\text{tr}\,\dd^\dagger\eta\mathring{\l}_{1,q-1}$ by using the identities \eqref{id2} and \eqref{so4}. Again, this expression is  zero.

Thus, the resulting equations of motion for the dual field $\widehat\o$ become
\be\label{correcteomsforo}
\text{tr}^{q}\,\dd^\dagger\dd\,\widehat{\o}_{[D-2,q]}=0
\ee
and we will now show their  equivalence to \eqref{eomcorrect}. Indeed,
\be\label{equation-correspondence}\begin{split}
&\text{tr}^{q}\,\dd^\dagger\dd\,\widehat{\o}=0\qquad\xLeftrightarrow{\eqref{id7}}\\
&\text{tr}\left(\widetilde\dd\text{tr}^q+(-1)^{q+1}\text{tr}^q\widetilde\dd\right)\dd\widehat\o=0\qquad\xLeftrightarrow{\eqref{id7}}\\
&(\dd^\dagger-\widetilde\dd\text{tr})\text{tr}^q\dd\widehat\o+(-1)^{q+1}\text{tr}^{q+1}\dd\widetilde\dd\widehat{\o}=0\qquad\xLeftrightarrow{\eqref{id1}}\\
&(-1)^q\text{tr}^q\dd^\dagger\dd\widehat\o+(-1)^{q+1}\text{tr}^{q+1}\dd\widetilde\dd\widehat{\o}=0\qquad\xLeftrightarrow{\eqref{correcteomsforo}}\qquad\text{tr}^{q+1}\dd\widetilde\dd\,\widehat{\o}=0\,,
\end{split}
\ee
where  we also used that $\widetilde\dd\text{tr}^{q+1}\dd\widehat{\o}_{[D-2,q]}=0$. The effect of the $q$ traces is that after complete gauge fixing, the number of independent field equation becomes simply $\binom{D-2}{D-q-2}$, which is equal to the ones for a $q$-form, as required.  
As explained in the main text, this proves Theorem \ref{theorem} for these values of parameters.

\paragraph{The fourth domain.} For the fourth domain $\{p=2, q\in[2,D-3]\}$, the parent Lagrangian reads as
\be\label{b14}
\mathcal{L}_{\text{P}}=\int_{\theta,\chi}F_{2,q}\star\mathcal{O}^{(2,q)}F_{2,q}+\int_{\theta,\chi}\dd F_{2,q}\ast\widetilde\ast\l_{3,q}
\ee
and the duality relation  after varying w.r.t. to $F$ is
\be\label{b15}
\star\mathcal{O}F=-\frac{1}{2}\widetilde\ast\dd\ast\l\,.
\ee
Using \eqref{starrelations} we can expand  $\star\,\mathcal{O}F=(-1)^{\epsilon}\ast\widetilde\ast(\mathcal{O}F-\eta\,\text{tr}\,\mathcal{O}F+\frac{1}{4}\,\eta^2\,\text{tr}^2\,\mathcal{O}F)$ and rewrite the above relation as
\be\label{b16}
\mathcal{O}\left(F-\eta\,\text{tr}\,F+\frac{1}{4}\,\eta^2\,\text{tr}^2\,F\right)=\frac{(-1)^{\epsilon+1}}{2}\,\dd^\dagger\l\,,
\ee
since $\mathcal{O}$ commutes both with $\eta$ and $\text{tr}$. Now one has to determine the inverse $\mathcal{O}^{-1}_{(2,q)}$ of $\mathcal{O}^{(2,q)}$. In the present case we have $p=2<q+1$ and we  find
\be\label{inverseo1}
\mathcal{O}^{-1}_{(2,q)}=b_1\,\mathbb{I}+b_2\,\s\,\widetilde \s+b_3\,\s^2\,\widetilde \s^2\,,
\ee
with coefficients given by
\be\label{b-coefficients}
b_1\equiv \frac{q+1}{q+2}\,,\qquad b_2\equiv \frac{q+1}{2(q+2)}\,,\qquad b_3\equiv-\frac{q+1}{2q(q+2)}\,.
\ee 
Acting with this $\mathcal{O}^{-1}$ on both sides of \eqref{b16} yields
\be\label{b19}
F-\eta\,\text{tr}F+\frac{1}{4}\,\eta^2\,\text{tr}^2F=\frac{(-1)^{\epsilon+1}}{2}\,\mathcal{O}^{-1}\dd^\dagger\l\,.
\ee 
One can then act with $\text{tr}^2$ on both sides of \eqref{b19} and repeatedly use the second relation in \eqref{so4} to find
\be\label{tr2F}
\text{tr}^2F=\frac{(-1)^{\epsilon+1}}{(D-q)(D-q-1)}\,\mathcal{O}^{-1}\text{tr}^2\,\dd^\dagger\l\,.
\ee
After substituting \eqref{tr2F} back into \eqref{b19}, one can then act with $\text{tr}$ on both sides of the resulting equation. Using again the second in \eqref{so4}, one finds 
\be\label{trF}
\text{tr}F=\frac{(-1)^{\epsilon}}{2(D-q-1)}\,\mathcal{O}^{-1}\left(\mathbb{I}-\frac{1}{D-q}\,\eta\,\text{tr}\right)\text{tr}\,\dd^\dagger\l\,.
\ee
Finally, plugging both \eqref{tr2F} and \eqref{trF} into \eqref{b19} yields
\be\label{Fdom4}
F=\frac{(-1)^{\epsilon+1}}{2}\,\mathcal{O}^{-1}\left(\mathbb{I}-\frac{1}{D-q-1}\,\eta\,\text{tr}-\frac{1}{2(D-q)(D-q-1)}\,\eta^2\,\text{tr}^2\right)\dd^\dagger\l\,.
\ee 
The dual Lagrangian in terms of $\l$ is then readily obtained by substituting \eqref{b15} and \eqref{Fdom4} into \eqref{b14}. In the standard spirit of the dualization procedure, one can then decompose the Lagrange multiplier into  traceless and trace parts
\be\label{decompositionfinal}
\l_{3,q}=\widehat\l_{3,q}+\eta\mathring{\l}_{2,q-1}
\ee
and define the $GL(D)$-irreducible dual field $\widehat{\o}_{[D-3,q]}$ to be the partial Hodge dual of the traceless part $\widehat{\l}_{3,q}$, i.e. $\widehat{\l}=\ast\,\widehat{\o}$. Not surprisingly, the $\mathring{\l}$-dependence in the dual Lagrangian cannot be algebraically canceled---c.f. Ref. \cite{Boulanger2}, where this is discussed for the special case of $q=D-3$. This is a common characteristic between the double standard dualization of generalized graviton fields and the exotic dualization of differential forms, since this was also the case in our previous analysis for the third domain of parameters. The logic remains the same, in that the Lagrangian contains additional off shell fields, whose elimination on shell  leads to equations of motion for the dual field only.

The equations of motion obtained by variation of the dual Lagrangian w.r.t. $\l$  read as
\be\label{eomsdualfinal}
\dd\mathcal{O}^{-1}\left(\mathbb{I}-\frac{1}{D-q-1}\,\eta\text{tr}-\frac{1}{2(D-q)(D-q-1)}\,\eta^2\text{tr}^2\right)\dd^\dagger\l=0\,.
\ee
Acting now on both sides with $\text{tr}^{q-1}\ast$ and using \eqref{identitiessigmatrace} and \eqref{id2}, the above equation can be rewritten as
\be\label{eomsdualfinal1}
\dd\s^{q-1}\,\mathcal{O}^{-1}\left(\mathbb{I}-\frac{1}{D-q-1}\,\eta\,\text{tr}-\frac{1}{2(D-q)(D-q-1)}\,\eta^2\,\text{tr}^2\right)\dd^\dagger\l=0\,.
\ee
Using the decomposition \eqref{decompositionfinal} and the map identities \eqref{so4}, \eqref{id7} and \eqref{id10}, we then compute
\be
\eta\,\text{tr}\,\dd^\dagger\l=-\eta\,\widetilde\dd\,\text{tr}\mathring{\l}-(D-q-1)\,\eta\,\dd^\dagger\mathring{\l}-\eta^2\,\text{tr}\,\dd^\dagger\mathring{\l}
\ee
and
\be\eta^2\,\text{tr}^2\,\dd^\dagger\l=\eta^2\,\text{tr}^2\,\widetilde\dd\mathring{\l}-2(D-q+1)\,\eta^2\,\text{tr}\,\dd^\dagger\mathring{\l}\,.
\ee
Plugging these expressions into \eqref{eomsdualfinal1} we obtain 
\be\label{eomsdualfinal2}
\dd\s^{q-1}\mathcal{O}^{-1}\dd^\dagger\hat\l+\dd\s^{q-1}\mathcal{O}^{-1}\widetilde\dd\mathring{\l}=0\,,
\ee
since, using  map identities from Appendix \ref{appa} and the inverse operator $\mathcal{O}^{-1}$ in \eqref{inverseo1},  all additional  terms
\be
\dd\s^{q-1}\mathcal{O}^{-1}\eta\,\widetilde\dd\text{tr}\mathring{\l}\,\,\,\,\,,\,\,\,\,\,\dd\s^{q-1}\mathcal{O}^{-1}\eta^2\text{tr}^2\widetilde\dd\mathring{\l}\qquad\text{and}\qquad \dd\s^{q-1}\mathcal{O}^{-1}\eta^2\text{tr}\dd^\dagger\mathring{\l}
\ee
vanish identically. Using now the identity \eqref{id3} and the nilpotency property of the exterior derivatives, one can write \eqref{eomsdualfinal2} as
\be\label{eomsdualfinal3}
\dd\s^{q-1}\mathcal{O}^{-1}\dd^\dagger\widehat\l=(-1)^q\dd\widetilde\dd\left(b_1\s^{q-1}\mathring{\l}+b_2\,\s^q\widetilde \s\mathring{\l}+b_3\,\s^{q+1}\widetilde \s^2\mathring{\l}\right)
\ee
for $b_1,b_2$ and $b_3$ given by \eqref{b-coefficients}. Then, we can use the first map identity \eqref{so4} successively to obtain 
\be
\s^q\,\widetilde \s\mathring{\l}=2q\,\s^{q-1}\mathring{\l}\qquad\text{and}\qquad \s^{q+1}\widetilde \s^2\mathring{\l}=2q(q+1)\s^{q-1}\mathring{\l}\,,
\ee
which can then be used to transform \eqref{eomsdualfinal3} into 
\be
\dd\s^{q-1}\mathcal{O}^{-1}\dd^\dagger\widehat\l=(-1)^q\dd\widetilde\dd\left(b_1+2qb_2+2q(q+1)b_3\right)\s^{q-1}\mathring{\l}\,.
\ee
However, the $b$-coefficients given by \eqref{b-coefficients} satisfy the algebraic equation
\be
b_1+2q\,b_2+2q(q+1)\,b_3=0
\ee
and therefore the $\mathring{\l}$-dependence in the equations of motion cancels completely. Thus we are left with
\be\label{eomsdualfinal4}
\dd\s^{q-1}\mathcal{O}^{-1}\dd^\dagger\widehat\l=0\,.
\ee 
As before, we can make a successive use of the first map identity in \eqref{so4} again to find 
\be
\s^q\,\widetilde \s\dd^\dagger\widehat{\l}=q\,\s^{q-1}\dd^\dagger\widehat\l\qquad\text{and}\qquad \s^{q+1}\widetilde \s^2\dd^\dagger\widehat\l=0\,.
\ee
Using these expressions, one can rewrite \eqref{eomsdualfinal4} as
\be
(b_1+q\,b_2)\,\dd\s^{q-1}\dd^\dagger\widehat\l=0\qquad\Leftrightarrow\qquad\dd\s^{q-1}\dd^\dagger\widehat\l=0\,.
\ee
These correspond exactly to the expected field equations 
\be \label{finalequationsofmotion}
\text{tr}^{q-1}\dd^\dagger\dd\widehat\o=0\qquad\Leftrightarrow\qquad\text{tr}^q\dd\widetilde\dd\widehat\o=0
\ee 
for the $GL(D)$-irreducible dual field $\widehat\o_{[D-3,q]}$ defined by $\widehat\l\equiv\ast\,\widehat\o$. Note that the above equation correspondence can be proven in the same way as  \eqref{equation-correspondence}.

Just as in the case of  exotic dualization of $q$-forms (third domain), here one can again define a Riemann-like tensor $R_{[D-2,q+1]}:= \dd\widetilde\dd\,\widehat\o_{[D-3,q]}$, as in the main text. The equations of motion $\text{tr}^{n}R=0$ will now imply the vanishing of the whole tensor $R$ for any $n<q$, thus the correct equations for the gauge field $\widehat\o$ will be $\text{tr}^{q}R=0$. These are exactly the equations \eqref{finalequationsofmotion} found above. Obviously, the effect of taking the $q$ traces is that the number of independent field equations for the dual field after complete gauge fixing is the same as the ones for the original irreducible field $\o_{[1,q]}$, as required.


\begin{thebibliography}{99}
	
	\bibitem{Hull}
	C.~M.~Hull,
	``Duality in gravity and higher spin gauge fields,''
	JHEP {\bf 0109} (2001) 027
	[hep-th/0107149].
	
	\bibitem{Bekaert:2002dt}
	X.~Bekaert and N.~Boulanger,
	``Tensor gauge fields in arbitrary representations of GL(D,R): Duality and Poincare lemma,''
	Commun.\ Math.\ Phys.\  {\bf 245} (2004) 27
	[hep-th/0208058].
	
	\bibitem{deMedeirosHull1}
	P.~de Medeiros and C.~Hull,
	``Exotic tensor gauge theory and duality,''
	Commun.\ Math.\ Phys.\  {\bf 235} (2003) 255
	[hep-th/0208155].
	
	
	\bibitem{Boulanger1}
	N.~Boulanger and D.~Ponomarev,
	``Frame-like off-shell dualization for mixed symmetry gauge fields,''
	J.\ Phys.\ A {\bf 46} (2013) 214014
	[arXiv:1206.2052 [hep-th]].
	
	\bibitem{Boulanger2}
	N.~Boulanger, P.~P.~Cook and D.~Ponomarev,
	``Off-Shell Hodge Dualities in Linearised Gravity and E11,''
	JHEP {\bf 1209} (2012) 089
	[arXiv:1205.2277 [hep-th]].
	
	
	\bibitem{DuboisViolette1}
	M.~Dubois-Violette and M.~Henneaux,
	``Generalized cohomology for irreducible tensor fields of mixed Young symmetry type,''
	Lett.\ Math.\ Phys.\  {\bf 49} (1999) 245
	[math/9907135].
	
	\bibitem{DuboisViolette2}
	M.~Dubois-Violette and M.~Henneaux,
	``Tensor fields of mixed Young symmetry type and N complexes,''
	Commun.\ Math.\ Phys.\  {\bf 226} (2002) 393
	[math/0110088 [math-qa]].
	
	\bibitem{Curtright:1980yk}
	T.~Curtright,
	``Generalized Gauge Fields,''
	Phys.\ Lett.\  {\bf 165B} (1985) 304.
	
	\bibitem{West:2001as}
	P.~C.~West,
	``E(11) and M theory,''
	Class.\ Quant.\ Grav.\  {\bf 18} (2001) 4443
	[hep-th/0104081].
	
	\bibitem{Cook:2004er}
	P.~P.~Cook and P.~C.~West,
	``G+++ and brane solutions,''
	Nucl.\ Phys.\ B {\bf 705} (2005) 111
	[hep-th/0405149].
	
	\bibitem{West:2004kb}
	P.~C.~West,
	``E(11) origin of brane charges and U-duality multiplets,''
	JHEP {\bf 0408} (2004) 052
	[hep-th/0406150].
	
	\bibitem{Bergshoeff:2011zk}
	E.~A.~Bergshoeff and F.~Riccioni,
	``String Solitons and T-duality,''
	JHEP {\bf 1105} (2011) 131
	[arXiv:1102.0934 [hep-th]].
	
	\bibitem{Bergshoeff:2010xc}
	E.~A.~Bergshoeff and F.~Riccioni,
	``D-Brane Wess-Zumino Terms and U-Duality,''
	JHEP {\bf 1011} (2010) 139
	[arXiv:1009.4657 [hep-th]].
	
	\bibitem{deBoer:2012ma}
	J.~de Boer and M.~Shigemori,
	``Exotic Branes in String Theory,''
	Phys.\ Rept.\  {\bf 532} (2013) 65
	[arXiv:1209.6056 [hep-th]].
	
	
	\bibitem{Chatzistavrakidis:2013jqa}
	A.~Chatzistavrakidis, F.~F.~Gautason, G.~Moutsopoulos and M.~Zagermann,
	``Effective actions of nongeometric five-branes,''
	Phys.\ Rev.\ D {\bf 89} (2014) no.6,  066004
	[arXiv:1309.2653 [hep-th]].
	
	\bibitem{Chatzistavrakidis:2014sua}
	A.~Chatzistavrakidis and F.~F.~Gautason,
	``U-dual branes and mixed symmetry tensor fields,''
	Fortsch.\ Phys.\  {\bf 62} (2014) 743
	[arXiv:1404.7635 [hep-th]].
	
	
	\bibitem{Bergshoeff:2015cba}
	E.~A.~Bergshoeff, V.~A.~Penas, F.~Riccioni and S.~Risoli,
	``Non-geometric fluxes and mixed-symmetry potentials,''
	JHEP {\bf 1511} (2015) 020
	[arXiv:1508.00780 [hep-th]].
	
	\bibitem{Bakhmatov:2017les}
	I.~Bakhmatov, D.~Berman, A.~Kleinschmidt, E.~Musaev and R.~Otsuki,
	``Exotic branes in Exceptional Field Theory: the SL(5) duality group,''
	JHEP {\bf 1808} (2018) 021
	[arXiv:1710.09740 [hep-th]].
	
	\bibitem{Fernandez-Melgarejo:2018yxq}
	J.~J.~Fern\'andez-Melgarejo, T.~Kimura and Y.~Sakatani,
	``Weaving the Exotic Web,''
	JHEP {\bf 1809} (2018) 072
	[arXiv:1805.12117 [hep-th]].
	
	\bibitem{Otsuki:2019owg}
	D.~S.~Berman, E.~T.~Musaev and R.~Otsuki,
	``Exotic Branes in M-Theory,''
	arXiv:1903.10247 [hep-th].
	
	\bibitem{Boulanger:2015mka}
	N.~Boulanger, P.~Sundell and P.~West,
	``Gauge fields and infinite chains of dualities,''
	JHEP {\bf 1509} (2015) 192
	[arXiv:1502.07909 [hep-th]].
	
	\bibitem{Boulanger2003}
	N.~Boulanger, S.~Cnockaert and M.~Henneaux,
	``A note on spin s duality,''
	JHEP {\bf 0306} (2003) 060
	[hep-th/0306023].
	
	\bibitem{Bergshoeff}
	E.~A.~Bergshoeff, O.~Hohm, V.~A.~Penas and F.~Riccioni,
	``Dual Double Field Theory,''
	JHEP {\bf 1606} (2016) 026
	[arXiv:1603.07380 [hep-th]].
	
	
	
	\bibitem{Bergshoeff:2016gub}
	E.~A.~Bergshoeff, O.~Hohm and F.~Riccioni,
	``Exotic Dual of Type II Double Field Theory,''
	Phys.\ Lett.\ B {\bf 767} (2017) 374
	[arXiv:1612.02691 [hep-th]].
	

	
	\bibitem{Severa2001}
	P.~\v{S}evera,
	``Some title containing the words `homotopy' and `symplectic',
	e.g. this one,''
	Travaux Math. {\bf 16} (2005) 121--137
	[arXiv:math.SG/0105080].
	
		\bibitem{dee1}
	D.~Roytenberg, 
	``On the structure of graded symplectic supermanifolds and
	Courant algebroids,''
	Contemp. Math. {\bf 315} (2002) 169--186
	[arXiv:math.SG/0203110].
	
	\bibitem{Qiu:2011qr}
	J.~Qiu and M.~Zabzine,
	``Introduction to Graded Geometry, Batalin-Vilkovisky Formalism and their Applications,''
	Archivum Math.\  {\bf 47} (2011) 143
	[arXiv:1105.2680 [math.QA]].
	
	\bibitem{Chatzistavrakidis1}
	A.~Chatzistavrakidis, F.~S.~Khoo, D.~Roest and P.~Schupp,
	``Tensor Galileons and Gravity,''
	JHEP {\bf 1703} (2017) 070
	[arXiv:1612.05991 [hep-th]].
	
	\bibitem{Bruce}
	A.~J.~Bruce and E.~Ibarguengoytia,
	``The Graded Differential Geometry of Mixed Symmetry Tensors,''
	Archivum Math.\ {\bf 55} (2019), No. 2
	[arXiv:1806.04048 [math-ph]].
	
	\bibitem{Bekaert:2003az}
	X.~Bekaert and N.~Boulanger,
	``On geometric equations and duality for free higher spins,''
	Phys. Lett. B \textbf{561} (2003), 183-190
	[arXiv:hep-th/0301243 [hep-th]].
	
	\bibitem{deMedeirosHull2}
	P.~de Medeiros and C.~Hull,
	``Geometric second order field equations for general tensor gauge fields,''
	JHEP {\bf 0305} (2003) 019
	[hep-th/0303036].
	
	\bibitem{deMedeiros:2003qel}
	P.~de Medeiros,
	``Massive gauge invariant field theories on spaces of constant curvature,''
	Class.\ Quant.\ Grav.\  {\bf 21} (2004) 2571
	doi:10.1088/0264-9381/21/11/004
	[hep-th/0311254].
	
	\bibitem{Nicolis:2008in}
	A.~Nicolis, R.~Rattazzi and E.~Trincherini,
	``The Galileon as a local modification of gravity,''
	Phys.\ Rev.\ D {\bf 79} (2009) 064036
	[arXiv:0811.2197 [hep-th]].
	
	\bibitem{Deffayet:2013lga}
	C.~Deffayet and D.~A.~Steer,
	``A formal introduction to Horndeski and Galileon theories and their generalizations,''
	Class.\ Quant.\ Grav.\  {\bf 30} (2013) 214006
	[arXiv:1307.2450 [hep-th]].
	
	\bibitem{Deffayet:2010zh}
	C.~Deffayet, S.~Deser and G.~Esposito-Farese,
	``Arbitrary $p$-form Galileons,''
	Phys.\ Rev.\ D {\bf 82} (2010) 061501
	[arXiv:1007.5278 [gr-qc]].
	

	
	\bibitem{Deffayet:2017eqq}
	C.~Deffayet, S.~Garcia-Saenz, S.~Mukohyama and V.~Sivanesan,
	``Classifying Galileon $p$-form theories,''
	Phys.\ Rev.\ D {\bf 96} (2017) no.4,  045014
	[arXiv:1704.02980 [hep-th]].
	
	\bibitem{Demessie}
	G.~A.~Demessie and C.~Saemann,
	``Higher Poincar\'e lemma and integrability,''
	J.\ Math.\ Phys.\  {\bf 56} (2015) no.8,  082902
	[arXiv:1406.5342 [hep-th]].
	
  
   \bibitem{prep} 
  A.~Chatzistavrakidis, G.~Karagiannis and P.~Schupp, in preparation
  
  \bibitem{WestL1}
  P.~C.~West,
  Class.\ Quant.\ Grav.\  {\bf 20} (2003) 2393
  [hep-th/0212291].
  
  \bibitem{WestL2}
  P.~West,
  arXiv:1411.0920 [hep-th].
  
  \bibitem{Polchinski:1995mt}
  J.~Polchinski,
  ``Dirichlet Branes and Ramond-Ramond charges,''
  Phys.\ Rev.\ Lett.\  {\bf 75} (1995) 4724
  [hep-th/9510017].

\bibitem{Romans:1985tz}
L.~J.~Romans,
``Massive N=2a Supergravity in Ten-Dimensions,''
Phys.\ Lett.\ B {\bf 169} (1986) 374
[Phys.\ Lett.\  {\bf 169B} (1986) 374].


  \bibitem{Bergshoeff:1996ui}
  E.~Bergshoeff, M.~de Roo, M.~B.~Green, G.~Papadopoulos and P.~K.~Townsend,
  ``Duality of type II 7 branes and 8 branes,''
  Nucl.\ Phys.\ B {\bf 470} (1996) 113
  [hep-th/9601150].
  
  \bibitem{Chamblin:1997nnu}
  A.~Chamblin and M.~J.~Perry,
  ``Dynamic D8-branes in IIA string theory,''
  hep-th/9712112.
  
  
  \bibitem{Bergshoeff:2012pm}
  E.~A.~Bergshoeff, A.~Kleinschmidt and F.~Riccioni,
  ``Supersymmetric Domain Walls,''
  Phys.\ Rev.\ D {\bf 86} (2012) 085043
  [arXiv:1206.5697 [hep-th]].
  
  \bibitem{Hassler:2013wsa}
  F.~Hassler and D.~L\"ust,
  ``Non-commutative/non-associative IIA (IIB) Q- and R-branes and their intersections,''
  JHEP {\bf 1307} (2013) 048
  [arXiv:1303.1413 [hep-th]].
  
  \bibitem{Bastianelli:2007pv}
  F.~Bastianelli, O.~Corradini and E.~Latini,
  JHEP {\bf 0702} (2007) 072
  [hep-th/0701055].
  
  \bibitem{Bonezzi:2018box}
  R.~Bonezzi, A.~Meyer and I.~Sachs,
  JHEP {\bf 1810} (2018) 025
  [arXiv:1807.07989 [hep-th]].
  

  
\end{thebibliography}
\end{document}